\documentstyle[prd,aps,epsfig,preprint,tighten]{revtex}
\begin{document}
\preprint{                                                   BARI-TH/259-97}
\draft
\title{        $\bar\nu_\mu\leftrightarrow\bar\nu_e$ mixing:
                    analysis of recent indications \\
        and implications for neutrino oscillation phenomenology}
\author{      G.\ L.\ Fogli, E.\ Lisi, and G.\ Scioscia}
\address{     Dipartimento di Fisica and Sezione INFN di Bari,\\
                  Via Amendola 173, I-70126 Bari, Italy}
\maketitle
                              \begin{abstract}
%...........................................................................
We reanalyze the recent data from the Liquid Scintillator Neutrino Detector 
(LSND) experiment, that might indicate $\bar\nu_\mu\leftrightarrow\bar\nu_e$ 
mixing. This indication is not completely excluded by the negative results 
of established accelerator and reactor neutrino oscillation searches. We 
quantify the region of compatibility by means of a thorough statistical 
analysis of all the available data, assuming both two-flavor and three-flavor
neutrino oscillations. The implications for various theoretical scenarios
and for future oscillation searches are studied. The relaxation of the LSND 
constraints under different assumptions in the statistical analysis is also 
investigated.
%...........................................................................
                               \end{abstract}

%%%%%%%%%%%%%%%%%%%%%%%%%%%%%%%%%%%%%%%%%%%%%%%%%%%%%%%%%%%%%%%%%%%%%%%%%%%%
\section{Introduction}
\label{sec:INT}
%%%%%%%%%%%%%%%%%%%%%%%%%%%%%%%%%%%%%%%%%%%%%%%%%%%%%%%%%%%%%%%%%%%%%%%%%%%%

	The Liquid Scintillator Neutrino Detector (LSND) \cite{At96b} 
collaboration at the Los Alamos Meson Physics Facility (LAMPF) claim 
evidence for neutrino oscillations $\bar\nu_\mu\leftrightarrow\bar\nu_e$  
from muon decay at rest \cite{At96a,At96c}, confirming  preliminary results 
reported in \cite{At95}.

	The neutrino energy range, $E_\nu\simeq20$--53 MeV, and the 
path length, $L\simeq30$~m, make the LSND experiment particularly 
sensitive to neutrino mass differences in the eV range, which is of 
great interest for cosmological structure formation (see, e.g., \cite{Pr95}). 
Mass differences  of the same order are probed by other accelerator
and reactor neutrino oscillation experiments \cite{PD96}, which, however, 
do not find evidence for oscillations. Therefore, the interpretation of 
the LSND results as a neutrino oscillation effect requires careful checks
and independent experimental confirmation. In the meanwhile, it is 
legitimate to study its possible implications in the phenomenology of 
neutrino masses and mixing.

	The usual and easiest way to study the LSND signal is in terms of 
pure two-flavor oscillations. More precisely, the tau neutrino is assumed to 
be decoupled, so that only $\nu_\mu\leftrightarrow\nu_e$ oscillations%
%------------------------
\footnote{In the following, neutrinos and antineutrinos will not be
distinguished. Their oscillations in vacuum are equivalent both in the
two-flavor limit and in the three-flavor scheme that we will use.}
%------------------------
may occur. The LSND oscillation signal can then be compared, in the
usual mass-mixing plane, to the limits coming from experiments probing
the same flavor appearance channel (E776 \cite{Bo92} and the 
Karlsruhe Rutherford Medium Energy Neutrino (KARMEN) experiment
\cite{Kl96,Kl97,Ei96}), as well as those probing $\nu_e$ disappearance 
(reactors G{\"o}sgen \cite{Za86}, Bugey \cite{Ac95}, and 
Krasnoyarsk \cite{Vi94}) and $\nu_\mu$ disappearance (CERN experiment 
CDHSW \cite{Dy84}). So far, this comparison has been simply performed by 
superposing the published 90\% C.L.\ limits (see, e.g., Fig.~31 in 
\cite{At96a}). However, the intersection of 90\% C.L.\ regions {\em is not\/} 
a 90\% C.L.\ region, and a more refined combination of the different 
experimental results is desirable.

	A three-flavor oscillation framework is required if, in addition, 
one wants to study the interplay between  LSND and the {\em tau\/} neutrino 
appearance experiments, such as  E531 at Fermilab \cite{Us86} (completed), 
the CERN Hybrid Oscillation Research Apparatus (CHORUS) 
\cite{Ma96}, and the Neutrino Oscillation Magnetic Detector
(NOMAD) experiment \cite{La96}. The most 
general three-flavor approach would imply a reanalysis of all the data in 
a huge parameter space. However, if two of the three neutrinos are assumed 
to be almost degenerate in mass (so that the oscillations are driven by just 
one mass parameter at the laboratory scale), then the three-flavor analysis 
becomes much more manageable (see \cite{Fo95,Mo94} and references therein). 
Many authors \cite{Mi95,Bi95,Bi96,Ba95,Ca96,Ta96,Ka96,Go97,To96,Ya96,Ac96}
have recently compared the LSND results with previous laboratory neutrino 
oscillation data within such approximation.  In all these analyses 
\cite{Mi95,Bi95,Bi96,Ba95,Ca96,Ta96,Ka96,Go97,To96,Ya96,Ac96}, the quoted 
limits on the neutrino masses and mixings were simply obtained as 
intersections of the region of parameter {\em preferred\/} by LSND and the 
regions {\em excluded\/} by the other oscillation experiments. As for the 
two-flavor case, no definite confidence level can be attached to such limits 
and to the corresponding scenarios, with the additional complication that 
the degrees of freedom increase when passing  from a two-flavor to a 
three-flavor oscillation scheme.

	Motivated by the wide interest in the LSND results, and by the lack 
of a rigorous comparison with the established accelerator and reactor 
neutrino oscillation limits, we set out to quantify this comparison in a 
statistically consistent approach. We build upon our previous work 
\cite{Fo95}, in which the results of the most constraining negative searches 
for neutrino oscillations were reanalyzed {\em ab initio\/}  and
combined in a global $\chi^2$ analysis. In the present work we include in 
the data set the recent LSND and KARMEN results, using both a two-flavor 
and a three-flavor approach. We obtain precise bounds on neutrino
mixings and mass differences, in the hypothesis that the LSND signal
can indeed be interpreted as evidence for neutrino oscillations. These
bounds are then used to discuss various scenarios described in 
\cite{Mi95,Bi95,Bi96,Ba95,Ca96,Ta96,Ka96,Go97,To96,Ya96,Ac96}, as well as 
to investigate possible expectations for future laboratory oscillation 
searches. As a byproduct of our reanalysis of the LSND data, we gain 
interesting information on the stability of the neutrino oscillation fit. 
In particular, we show how the LSND constraints get weakened
when the energy distribution of the LSND signal is gradually integrated out.

	Our work is organized as follows. In Sec.~II we combine the LSND
results with all the other established accelerator and reactor 
data relevant for two-flavor oscillations $\nu_\mu\leftrightarrow\nu_e$, 
and derive bounds on the neutrino mass and mixing parameters. In Sec.~III 
we include also the  data from the established $\nu_\tau$ appearance 
searches in a three-flavor oscillation framework with hierarchical mass 
differences. In Sec.~IV we investigate the implications of our results on 
some phenomenological scenarios and on future oscillation searches.
We draw our conclusions in Sec.~V. Our reanalysis of the LSND data is 
presented in Appendix A, together with a critical investigation of the 
stability of the fit. The reanalysis of the KARMEN data is detailed in 
Appendix B.

%%%%%%%%%%%%%%%%%%%%%%%%%%%%%%%%%%%%%%%%%%%%%%%%%%%%%%%%%%%%%%%%%%%%%%%%%%%%
\section{Analysis in two flavors}
\label{sec:TWO}
%%%%%%%%%%%%%%%%%%%%%%%%%%%%%%%%%%%%%%%%%%%%%%%%%%%%%%%%%%%%%%%%%%%%%%%%%%%%

	Let us assume pure two-flavor oscillations between the electron and 
muon neutrino, with mass-mixing parameters $(m^2,\,\sin^2 2\theta)$. The 
event excess claimed by the LSND collaboration constrains these parameters 
in a relatively narrow region, as shown in Fig.~31 of \cite{At96a}.

	We have independently reanalyzed the published LSND data with a 
maximum likelihood method,  as detailed in Appendix~A. The results are shown
in  Fig.~\ref{F:1}. The 90 and 99 \% C.L.\ contours compare well  with the 
LSND confidence limits (Fig.~31 of \cite{At96a}), except for a ``wiggle'' 
at $m^2\simeq 3$ eV$^2$ that, in our fit, extends  to higher values of 
$\sin^2 2\theta$. This difference is irrelevant, being  located in the 
region excluded by  other established neutrino oscillation searches (as we 
shall see). Our best fit is reached for 
$(m^2,\,\sin^2 2\theta)=(3{\rm\ eV}^2,\,0.009)$. The best-fit point is not 
particularly important in itself, since good fits are also reached in other 
zones that form a set of  almost-degenerate  maxima of the likelihood function 
$\cal L$, as can be realized by looking at the 68\% C.L. contours of Fig.~1. 
One of the local maxima practically coincides with the best-fit point of the 
published LSND  analysis \cite{At96a} 
$(m^2,\,\sin^2 2\theta)=(19{\rm\ eV}^2,0.006)$. 
The fit disfavors values of $m^2$ close to integer multiples of 4.3 eV$^2$, 
for which the oscillation probability of the high-energy neutrinos
(which are responsible for most of the event excess) is suppressed.

	The reader is referred to Appendix~A for an extensive discussion
of the LSND data fit. Here we just anticipate that  variations in the 
statistical treatment of  the data  may produce nonnegligible changes in 
the fit. In particular, the evidence in favor of neutrino oscillations 
appears to be weakened as the energy distribution of the LSND event excess 
is gradually integrated. Therefore, we think that the exact shape of the 
present LSND confidence contours should be taken with a grain of salt, 
while waiting for higher statistics and more stable fits.  For definiteness,
however, we will always refer to the LSND limits as derived from our maximum 
likelihood reanalysis (Fig.~\ref{F:1}) of the published LSND data.
The combination with the data from other experiments is
performed through the $\chi^2$ method. To this purpose, the LSND likelihood 
is transformed as $\chi^2=-\ln {\cal L}/2$; the LSND $\chi^2$ is then summed 
to the $\chi^2$'s of the other experiments.

	The LSND indications can be directly compared with the (negative) 
results from two experiments that probed the same oscillation channel 
$\nu_\mu\leftrightarrow \nu_e$, namely KARMEN \cite{Kl97} and E776 
\cite{Bo92}. Our reanalysis of the KARMEN data is described in  detail in 
Appendix~B. The results agree very well with the published oscillation 
bounds \cite{Kl97}. Figure~\ref{F:2}(a) shows our estimated 90 and 99 \%
C.L.\ contours for the KARMEN experiment ($\Delta\chi^2=4.61$ and $9.21$ 
for two degrees of freedom).

	Figure~\ref{F:2}(b) shows the results of a combined fit to the LSND 
and KARMEN data. The regions allowed at 90 and 99 \% C.L.\ are  only
slightly restricted with respect to Fig.~\ref{F:1}, as most of the LSND 
oscillation signal is beyond the present KARMEN sensitivity. However, the 
KARMEN experiment should  be able to test a significantly larger region of 
the parameter space in about two years \cite{Kl97}.

	The results from the E776 experiment \cite{Bo92} are more constraining
than those from KARMEN. In Fig.~\ref{F:2}(c) we report the E776 exclusion 
contours as derived by our reanalysis of the E776 data \cite{Fo95}.
The combination with the LSND and KARMEN data is shown in  Fig.~\ref{F:2}(d).
Notice that the 90\% C.L.\ region of Fig.~\ref{F:2}(d) is significantly
larger than the  intersection (not shown) of the 90\% C.L.\ regions allowed 
by the LSND, KARMEN, and E776 experiments separately, especially for small 
mixing. The reason is that the E776 (and KARMEN) negative results are 
compatible with no mixing, and thus  push the combined fit towards
values of the mixing angle smaller than those allowed by LSND alone. 
Moreover, the addition of the E776 data weakens the global  evidence for 
oscillations, so that the no oscillation limit is allowed at 99\% C.L.\ 
in Fig.~\ref{F:2}(d), while it is excluded in  Fig.~\ref{F:2}(b). In other 
words,  by adding the E776 and KARMEN data, the gradient of the
$\chi^2$ is suppressed for small values of the oscillation parameters, 
so that the 90\% C.L.\ constraints are significantly relaxed in that
direction, and the 99\% C.L.\ left contour disappears. These results show 
that a combined analysis of the neutrino oscillation data may give  rather 
different bounds than the naive ``superposition'' of oscillation limits, 
especially when the individual experimental results pull the fit in 
different directions.

	In the two-flavor approximation, the $\nu_e$ disappearance searches 
at reactor experiments can be interpreted as  probes of $\nu_e\leftrightarrow
\nu_\mu$ oscillations. The data from the G{\"o}sgen \cite{Za86}, Bugey 
\cite{Ac95}, and Krasnoyarsk \cite{Vi94} experiments were reanalyzed and 
combined in \cite{Fo95}. The results are shown in Fig.~\ref{F:2}(e). 
When combined with the LSND, KARMEN, and E776 data, the reactor
results exclude the region at large mixing and low mass differences, as 
shown in Fig.~\ref{F:2}(f). Notice that the  narrow, disconnected regions 
allowed at 90\% C.L.\ at large $m^2$ in Fig.~\ref{F:2}(f) would disappear 
if a simple superposition of oscillation plots were made instead of a global 
fit.

	Finally, we mention that the results from $\nu_\mu$ disappearance 
searches can also be added straightforwardly in the two-flavor approximation. 
However, the available data (CDHSW experiment \cite{Dy84})  probe relatively 
large mixing angles, and we have checked that their addition produces 
negligible changes (not shown) to the contours of Fig.~\ref{F:2}(f). The 
CDHSW data are anyway added, in the way described in \cite{Fo95}, in the 
global three-flavor analysis of the following section.

%%%%%%%%%%%%%%%%%%%%%%%%%%%%%%%%%%%%%%%%%%%%%%%%%%%%%%%%%%%%%%%%%%%%%%%%%%%%
\section{Analysis in three flavors}
\label{sec:THR}
%%%%%%%%%%%%%%%%%%%%%%%%%%%%%%%%%%%%%%%%%%%%%%%%%%%%%%%%%%%%%%%%%%%%%%%%%%%%

	In this section we combine with the $\chi^2$ method the data from 
eight (reanalyzed) neutrino oscillation experiments (LSND, KARMEN, E776, 
G{\"o}sgen, Krasnoyarsk, Bugey, CDHSW, and E531 {\cite{E531}), working 
in the simple three-flavor framework described in \cite{Fo95}. This 
framework is characterized by the assumption that two out of three neutrino 
mass eigenstates $\nu_i\;(i=1,2,3)$ are effectively degenerate in mass, 
$|m^2_2-m^2_1|\simeq 0$, while the mass gap of the ``lone'' state $\nu_3$ 
is taken in the range of laboratory neutrino oscillation experiments: 
$m^2=|m^2_3-m^2_{2,1}| \gtrsim 10^{-3}$ eV$^2$. The neutrino oscillation 
parameter space is then spanned by three variables only: the dominant 
square mass difference $m^2$, and the mixing angles $\phi$ and $\psi$ 
(equal to $\theta_{13}$ and $\theta_{23}$ in the standard ordering of mixing 
matrices \cite{PD96}). More precisely:
%...........................................................................
\begin{mathletters}
\begin{equation}
 m^2  =  |m^2_3-m^2_{2,1}| \gg |m^2_2-m^2_1|              \ ,
\end{equation}
\begin{equation}
\nu_3 = \sin\phi\;\nu_e+\cos\phi\,(\sin\psi\;\nu_\mu+\cos\psi\;\nu_\tau)\ ,
\end{equation}
\end{mathletters}
%...........................................................................
with $\psi,\,\phi\in[0,\,\pi/2]$. We prefer to use the variables 
$(\tan^2\psi,\,\tan^2\phi)$, which prove very useful for graphical 
representations \cite{Fo95}. Notice that the two-flavor oscillation limits 
$\nu_\mu\leftrightarrow\nu_\tau$, $\nu_e\leftrightarrow\nu_\tau$, and
$\nu_e\leftrightarrow\nu_\mu$, are reached for $\phi=0$, $\psi=0$, and 
$\psi=\pi/2$, respectively. The reader is referred to \cite{Fo95} for 
further details and bibliography.

	Figure~\ref{F:3} shows the results of our analysis in the 
$(m^2,\,\tan^2\phi)$ plane, for twelve representative values of 
$\tan^2\psi$. The 90 and 99 \% C.L.\ contours shown in each panel represent 
sections (at fixed $\tan^2\psi$) of the confidence volume defined by
$\chi^2-\chi^2_{\rm min}=6.25$ and $11.36$ respectively (for three degrees 
of freedom).

	In the first panel of Fig.~\ref{F:3} the value of $\tan^2\psi$
is very large, corresponding to almost pure $\nu_\mu\leftrightarrow\nu_e$
oscillations $(\psi\to\pi/2)$. In fact, the limits close to the left
side of the first panel represent a mapping of the two-flavor oscillation 
limits shown in Fig.~\ref{F:2}(f) (modulo the different number of degrees of 
freedom). The mirror limits on the right $(\phi\to\pi/2-\phi)$ in the same 
panel come from enlarging the mixing angle range from $[0,\,\pi/4]$ 
(2 flavors) to $[0,\,\pi/2]$ (3 flavors or more).

	In the subsequent panels the value of $\tan^2\psi$ gradually 
decreases and the mixing of $\nu_3$ with $\nu_\tau$ increases. The $2\nu$ 
left-right symmetry of the plots $(\phi\to\pi/2-\phi)$ is broken. 
The allowed regions get gradually shrinked, since for $\psi\to 0$
one approaches  the two-flavor oscillation limit 
$\nu_e\leftrightarrow\nu_\tau$, which is not compatible  with the LSND data. 
The allowed regions on the left half of each panel disappear more rapidly, 
since for $\psi\to 0$ {\em and\/} $\phi\to 0$ the relevant mass eigenstate 
$(\nu_3)$ tends to become a flavor eigenstate ($\nu_\tau$) and the oscillation
phenomenon vanishes. Notice that  the combination of all the data 
constrains $m^2$ above $\sim 0.2$ eV$^2$ at 90\% C.L. However,
at 99\% C.L. there is no lower bound on $m^2$ and the global
fit becomes compatible with no oscillations.

	Figure~\ref{F:4} shows an alternative representation of the same
confidence volume of Fig.~\ref{F:3}. The volume is shown  in 
$(\tan^2\psi,\,\tan^2\phi)$ sections at fixed values of $m^2$. We remind 
that the left, right, and lower side of each panel correspond,
asymptotically, to pure  $\nu_e\leftrightarrow\nu_\tau$, 
$\nu_e\leftrightarrow\nu_\mu$, and
$\nu_\mu\leftrightarrow\nu_\tau$ oscillations, respectively.
In Fig.~\ref{F:4}, the region allowed at 90\% C.L.\ generally
consists of two disconnected parts at low and high $\phi$.
These zones would be continuously connected by a $\subset$-shaped band 
if only the LSND data were fitted \cite{Fo95}. However, the data from 
all other oscillation experiments strongly disfavor the central
part of each panel, corresponding to large (and not observed) three-flavor 
mixing. The width of the 90\% allowed regions is very sensitive to small 
variations of $m^2$ around multiples of 4.3 eV$^2$ where, as noticed, the 
LSND oscillation probability is suppressed.  This is particularly evident
in the panel at $m^2=13$ eV$^2$ where no combination of mixing angles
is allowed at 90\% C.L.  As in Fig.~\ref{F:3}, the no oscillation scenario 
(corresponding to the lower corners and the upper side of each panel) 
appears to be allowed at 99\% C.L.\ by the combination of all the data.

 	Figures~\ref{F:3} and \ref{F:4} are the main result of this work.
They represent a concise summary (in a three-flavor scheme) of the most 
constraining  neutrino oscillation data available, in the hypothesis that 
the LSND signal can indeed be interpreted as a signal of 
neutrino oscillations.

%%%%%%%%%%%%%%%%%%%%%%%%%%%%%%%%%%%%%%%%%%%%%%%%%%%%%%%%%%%%%%%%%%%%%%%%%%%%
\section{Phenomenological implications}
\label{sec:IMP}
%%%%%%%%%%%%%%%%%%%%%%%%%%%%%%%%%%%%%%%%%%%%%%%%%%%%%%%%%%%%%%%%%%%%%%%%%%%%

	In this section we study the implication of the results
shown in Figs.~\ref{F:3} and \ref{F:4} for various theoretical scenarios
and for short-baseline neutrino oscillation experiments.

%%%%%%%%%%%%%%%%%%%%%%%%%%%%%%%%%%%%%%%%%%%%%%%%%%%%%%%%%%%%%%%%%%%%%%%%%%%%
\subsection{Implications for theoretical scenarios}

	Apart from the LSND indications, there are (older) hints of
neutrino oscillations from the solar neutrino problem and from
the atmospheric neutrino anomaly. These additional data have
been recently analyzed, within the same three-flavor framework adopted
in this work, in \cite{Fo96} for solar neutrinos and in \cite{Fo97}
for atmospheric neutrinos.

	The results obtained in this work and in \cite{Fo96,Fo97} indicate 
that the LSND and atmospheric data are difficult to reconcile at 90\% C.L.\ 
if the subdominant neutrino mass difference is called to solve the solar
neutrino problem \cite{Fo96}. In fact, from the LSND analysis 
(Figs.~\ref{F:3} and \ref{F:4}) we derive $m^2 \gtrsim 0.22$ eV$^2$, while 
from the  atmospheric neutrino analysis \cite{Fo97} we derive the 
incompatible bound $m^2 \lesssim 0.1$ eV$^2$. These data could be  marginally 
reconciled only by dropping the information provided by the multi-GeV event 
sample of the Kamiokande experiment, as observed in \cite{Ca96} 
(see also \cite{Ya96}).  If no data are excluded, however, one cannot 
achieve with three-flavor oscillations a good fit of the solar, atmospheric, 
and LSND data  {\em at the same time\/}. The addition of a fourth, sterile
neutrino might reconcile all the data without exclusions 
(see, e.g., \cite{Go97}), but we think that more robust experimental checks 
must be performed before adopting such an extreme interpretation.

	Let us consider now the interplay between the LSND and the solar
neutrino data. As shown in Fig.~\ref{F:4}, at any given $m^2$ there are two 
regions allowed at 90\% C.L.\ at small and large $\phi$. However, only the 
small $\phi$ region survives the comparison with solar neutrino data 
\cite{Fo96}, since at large $\phi$ the solar neutrino deficit becomes
almost energy-independent, contrary to the present experimental
evidence \cite{KrPe}. The small $\phi$, large $\psi$  allowed solution of 
Fig.~\ref{F:4} would indicate that   the ``lone'' neutrino state
$\nu_3$ is dominantly a $\nu_\mu$, thus excluding a simultaneous
hierarchy of masses and mixings.

	Many semi-quantitative three-flavor analyses of laboratory neutrino 
data  cornered a third, marginal ``90\% LSND allowed region'' at small 
values of $\phi$ and $\psi$ \cite{Mi95,Ba95,Ta96,Ka96,Ac96}, corresponding 
to  $\nu_3$ dominantly coupled with $\nu_\tau$. This possibility is 
appealing since small mixings  are generally regarded as more natural. 
However,  our quantitative analysis shows that  such region does not appear 
at 90\% C.L.\ for any value $m^2$ and, therefore, it is unlikely that
the $\nu_\tau$ is the dominant flavor component of $\nu_3$.

	The threefold maximal mixing scenario (see, e.g., \cite{Ha95}),
corresponding to $(\tan^2\psi,\,\tan^2\phi)=(1,\,1/2)$, is  excluded at more 
than $99\%$ C.L. for all values of $m^2$ in the LSND sensitivity range, as 
can be seen in Fig.~\ref{F:4}. The quasi-maximal mixing scenario proposed
in \cite{To96}, corresponding to $\psi\simeq \pi/4$ and small $\phi$,
also appears to be strongly disfavored by our analysis.

	In conclusion, if one believes in the indications from solar 
{\em and\/} LSND neutrino experiments, then the dominant flavor component 
of  the mass eigenstate $\nu_3$ is the muon neutrino. This conclusion
is relevant for model building. The precise bounds on the
$\nu_3$ flavor content can be derived from  Fig.~\ref{F:4} at any given 
value of $m^2$, taking into account that the ``large $\phi$'' solution is 
excluded by the analysis of solar neutrino data \cite{Fo96}.

%%%%%%%%%%%%%%%%%%%%%%%%%%%%%%%%%%%%%%%%%%%%%%%%%%%%%%%%%%%%%%%%%%%%%%%%%%%%
\subsection{Implications for short-baseline experiments}

	The implications of the LSND results for experiments  probing the 
same oscillation channel $(\nu_\mu\leftrightarrow\nu_e)$ are straightforward: 
they should observe a positive indication for neutrino oscillation, or 
disprove it, if they reach (at least) the LSND sensitivity. This goal 
should be reached by the KARMEN experiment in about two years \cite{Kl97}.
We mention in passing that the NOMAD experiment is also sensitive to 
$\nu_e$ appearance for $m^2\gtrsim 10$~eV$^2$, where it is expected to 
improve the existing limits \cite{DiLe}.

	The implications for experiments probing different oscillation 
channels are less evident and require a three-flavor language. Here we 
focus on $\nu_\mu\leftrightarrow\nu_\tau$ searches, currently being 
performed by the CHORUS \cite{Ma96} and NOMAD \cite{La96} experiments at 
CERN. Both experiments are expected to release soon the preliminary results 
of the first two-year run (1994--95), and are scheduled for additional two 
years  of data taking (1996--97) \cite{St97}. Two proposals for their 
upgrade are  currently being considered at CERN, namely, the
Tracking and Emulsion for Neutrino Oscillation Research (TENOR)
\cite{Er96}, and the Neutrino Apparatus with Improved Capabilities
(NAUSICAA) \cite{GC96}. The Cosmologically Significant Mass Oscillation
Search (COSMOS) experiment \cite{Si96} at Fermilab is also planned to probe 
the $\nu_\mu\leftrightarrow\nu_\tau$
channel with an expected sensitivity greater than CHORUS and NOMAD but
somewhat lower than TENOR or NAUSICAA.

	Figure~\ref{F:5} shows the 90\% C.L.\ regions that can be probed by
CHORUS or NOMAD (in two years \cite{Ma96,La96} and four years 
\cite{DiLe,St97}),  COSMOS, and TENOR or NAUSICAA, using 
the $(\tan^2\psi,\,\tan^2\phi)$ representation at fixed values of $m^2$. 
Also shown is the region preferred at 90\% C.L.\ by the combination of all 
available data (including LSND), as taken from Fig.~\ref{F:4}. It can be 
seen that the CHORUS and NOMAD experiments cannot probe, with a
two-year statistics,  the region preferred by all the data at 90\% C.L.\
(except for a marginal zone at $m^2\simeq10$ eV$^2$),
as already noticed in \cite{Fo95}.  However, it is interesting to note that 
two additional years of data taking will make the CHORUS and NOMAD
experiments able to probe a fraction of the small $\phi$, large $\psi$ 
allowed region (see, e.g., the two top panels in Fig.~\ref{F:5}).
The COSMOS experiment and the TENOR or NAUSICAA projects will not only 
improve the sensitivity to the small $\phi$ solution at low $m^2$, but 
could even probe the large $\phi$ solution at large $m^2$ (as in the 
first panel of Fig.~\ref{F:5}).

	It follows from Fig.~\ref{F:5} that, if an oscillation 
signal shows up in CHORUS or NOMAD (after a 4-year run)  and if the LSND 
claim is confirmed, then the mixing angles will be  tightly constrained in 
the small $\phi$, large $\psi$ region. Such a solution would be  compatible 
with solar neutrinos \cite{Fo96}, but not with atmospheric neutrinos 
\cite{Fo97}.  Conversely, if the CHORUS and NOMAD oscillation
searches give negative results, then the present small $\phi$,
large $\psi$ solution will be strongly reduced. The proposed COSMOS, TENOR, 
and NAUSICAA experiment might probe, in part, the additional possibility 
offered by the large  $\phi$ solution. We conclude that the running and future 
$\nu_\mu\leftrightarrow\nu_\tau$ experiments can explore an interesting
fraction of the LSND allowed region, and will provide in any case
decisive constraints on the neutrino mixing angles.

	We finally mention that the $\nu_e\to\nu_e$ reactor experiments 
in construction at Chooz \cite{Choo} and Palo Verde \cite{Palo} are planned
to improve the existing limits in the range 
$10^{-3} {\rm\ eV}^2 \lesssim m^2\lesssim 10^{-2} {\rm\ eV}^2$. However,
they  are not expected to improve the available Bugey bounds in the range of
$m^2$ relevant for LSND ($m^2\gtrsim 0.22$ eV$^2$) and, therefore,
should not have enough sensitivity  to probe
the 90\% C.L.\ regions shown in Fig.~\ref{F:4}. More precisely,
the Chooz or Palo Verde  sensitivity region
would appear as a horizontal band
in the $(\tan^2\psi,\,\tan^2\phi)$ plane \cite{Fo95}, with an expected
width given by
$0.04 \lesssim \tan^2\phi \lesssim 25$, which does not overlap with 
the 90\% C.L.\ allowed regions of Fig.~\ref{F:4}.

%%%%%%%%%%%%%%%%%%%%%%%%%%%%%%%%%%%%%%%%%%%%%%%%%%%%%%%%%%%%%%%%%%%%%%%%%%%%
\section{Conclusions}
\label{sec:SUM}
%%%%%%%%%%%%%%%%%%%%%%%%%%%%%%%%%%%%%%%%%%%%%%%%%%%%%%%%%%%%%%%%%%%%%%%%%%%%

	We have studied the interplay between the recent LSND indications 
and the negative results of the established oscillation searches at 
accelerators  (E776, KARMEN, E531, CDHSW) and reactors (G{\"o}sgen, Bugey, 
Krasnoyarsk). A thorough analysis has been performed assuming two-flavor 
and three-flavor mixing. The region of neutrino oscillation parameters 
preferred by all the data  has been determined. In particular, the 
three-flavor analysis in the $(m^2,\,\psi\,\,\phi)$ space typically
favors two solutions, one at small and the other 
at large $\phi$. Only the small $\phi$ 
solution is consistent with the solar neutrino data. A small fraction of the 
small $\phi$ solution  appears to be in the range explorable  by the CHORUS 
and NOMAD experiments (four year run).  A larger fraction of this 
solution, and even a small part of the large $\phi$ region, could be 
probed by the COSMOS experiment at Fermilab  and (with greater sensitivity) 
by the TENOR or NAUSICAA experiments  at CERN. Finally, it has been shown 
(Appendix~A) how different statistical treatments of the  published 
LSND data  affect the bounds on the neutrino oscillation parameters.

%%%%%%%%%%%%%%%%%%%%%%%%%%%%%%%%%%%%%%%%%%%%%%%%%%%%%%%%%%%%%%%%%%%%%%%%%%%%
\acknowledgments
%%%%%%%%%%%%%%%%%%%%%%%%%%%%%%%%%%%%%%%%%%%%%%%%%%%%%%%%%%%%%%%%%%%%%%%%%%%%

We acknowledge useful discussions with L.\ DiLella, U.\ Dore, J.\ Ellis,
and P.\ Strolin. We thank the organizers of the CERN
Joint Neutrino Meeting of the CHORUS and NOMAD Collaboration, where
preliminary results of this work were presented.
One of us (G.S.) would like to thank the CERN Particle Physics 
Experiments Division and the CHORUS Collaboration for kind hospitality 
during  completion of this work, and the INFN for partial support. 
This work was performed under the auspices of the Theoretical and 
Astroparticle Network (TAN) of the E.E.C.

\appendix
%%%%%%%%%%%%%%%%%%%%%%%%%%%%%%%%%%%%%%%%%%%%%%%%%%%%%%%%%%%%%%%%%%%%%%%%%%%%
\section{Reanalysis of the LSND data}
\label{app:LSN}
%%%%%%%%%%%%%%%%%%%%%%%%%%%%%%%%%%%%%%%%%%%%%%%%%%%%%%%%%%%%%%%%%%%%%%%%%%%%

	The Liquid Scintillator Neutrino Detector (LSND) experiment 
\cite{At96b} searches for $\bar\nu_e$ appearance in a $\bar\nu_\mu$ beam 
from the decay (at rest)
%...........................................................................
\begin{eqnarray}
\label{eq:DAR}
\pi^+ & \to & \mu^+ + \nu_\mu \nonumber\\
      &     & \hookrightarrow e^+ + \nu_e + \bar\nu_\mu\ ,
\end{eqnarray}
%...........................................................................
through the reaction 
%...........................................................................
\begin{equation}
\label{eq:POS}
\bar\nu_e+p\to e^+ + n\ .
\end{equation}
%...........................................................................

	The maximum $\bar\nu_\mu$ energy is $E_\nu^{\text{max}}=52.8$ MeV.
For any candidate positron, the LSND detector  can determine both its energy 
$E_e$ and its vertex coordinate $L$ (equal to the neutrino path length).  
The detector covers the range $L=29.8\pm3.8$~m.

	Positron-like events have been observed \cite{At96a,At96c} in 
excess of the estimated background. If this excess is interpreted as a 
signal of neutrino oscillations, a region of (two-flavor) neutrino mass 
and mixing parameters appears to be preferred by the data analysis, as 
shown in Fig.~31 of  \cite{At96a}.

	In this Appendix we describe how  such region can be reproduced to 
a good accuracy by reanalyzing the published LSND data, and how it gets 
modified by changing some assumptions.  We fix the terminology
in Sec.~A1,  develop the standard analysis in Sec.~A2, and discuss 
variations of the standard analysis in Sec.~A3.

%%%%%%%%%%%%%%%%%%%%%%%%%%%%%%%%%%%%%%%%%%%%%%%%%%%%%%%%%%%%%%%%%%%%%%%%%%%%
\subsection{Basic definitions}

	The total LSND signal $S$  is defined as the sum of the total 
(beam-off  plus beam-related) background $B$ and of the event excess $E$ 
due to  possible oscillations,
%...........................................................................
\begin{equation}
S = B + E\ .
\end{equation}
%...........................................................................

	The subscript ``exp'' and ``theo'' will be  used to distinguish the 
observed values of $S$ and $E$ from  the theoretical  estimates, respectively. 
No subscript is attached to $B$, which is always a simulated quantity.

	For a given experimental distribution  of the signal (divided in $N$ 
bins), $\{n_i\}_{i=1\dots N}$,  and the associated theoretical distribution 
$\{\mu_i\}_{i=1\dots N}$,  a likelihood function $\cal L$ can be defined as
%............................................................................
\begin{equation}
{\cal L} = \displaystyle
\prod_{i=1}^N \frac{1}{n_i!}\,{\mu_i}^{n_i}\,e^{-\mu_i}\ .
\end{equation}
%............................................................................

	The analysis performed by the LSND collaboration \cite{At96a} is 
essentially based on the likelihood  ${\cal L}_{EL}$
associated to the (experimental and theoretical) double
differential distribution $d^2S/dE_e\,dL$,
%.............................................................................
\begin{equation}
\label{eq:Like}
{\cal L}_{EL} = {\cal L}(d^2S/dE_e\,dL) \ ,
\end{equation}
%.............................................................................
having divided  the positron energy 
range $(20{\rm\ MeV}\leq E_e \leq 60{\rm\ MeV})$ into 20 bins
and the length range $(26.0 {\rm\ m}\leq L \leq 33.6 {\rm\ m})$
into 38 bins.%
%------------------------------
\footnote{Actually, the complete LSND analysis \protect\cite{At96a}
includes also subdominant information related to the positron
scattering angle and  the event pattern recognition.}
%------------------------------

	Unfortunately, the double differential distributions 
$d^2S_X/dE_e\,dL$ ($X={\rm exp,}\,{\rm theo}$) are not published in 
Refs.\ \cite{At96b,At96a}, so that the likelihood function
in Eq.~(\ref{eq:Like}) cannot be reproduced exactly.
However, one can recover at least the projected distributions 
$dS_X/dE_e$ and $dS_X/dL$  $(X={\rm exp},\,{\rm theo})$ 
from \cite{At96b,At96a}  and thus calculate the corresponding 
likelihoods ${\cal L}_E$ and ${\cal L}_L$.  In the absence of more 
detailed information, we assume that the total likelihood is factorizable,
%.............................................................................
\begin{eqnarray}
{\cal L}_{EL} &\sim& {\cal L}_E \times {\cal L}_L \nonumber\\
&=&{\cal L}(dS/dE_e) \times {\cal L}(dS/dL)\ ,
\label{eq:Fact}
\end{eqnarray}
%.............................................................................
as if $dS/dE_e$ and $dS/dL$ were independent. Although such statistical
independence is never exactly realized, we will see
{\em a posteriori\/} that  the published LSND oscillation analysis 
\cite{At96a} is  reproduced rather well through the approximation 
Eq.~(\ref{eq:Fact}).

	In the next subsection we discuss how the relevant distributions 
of the signal can be recovered from Refs.~\cite{At96b,At96a} and then 
used to perform a maximum likelihood analysis.

%%%%%%%%%%%%%%%%%%%%%%%%%%%%%%%%%%%%%%%%%%%%%%%%%%%%%%%%%%%%%%%%%%%%%%%%%%%%
\subsection{Standard Analysis}

	The approximate likelihood function in Eq.~(\ref{eq:Fact})
is based on the four distributions $dS_{\rm exp}/dE_e$, $dS_{\rm exp}/dL$,
$dS_{\rm theo}/dE_e$, and $dS_{\rm theo}/dL$.  The first two can be read 
off Fig.~30(a) and 30(d) of \cite{At96a} respectively
(dots with statistical error bars). The last two are estimated as follows.

	The double differential (theoretical) distribution of events $E$ in 
excess of the background is calculated as 
%...........................................................................
\begin{equation}
\frac{d^2 E_{\rm theo}}{dE_e\,dL}=
{\cal N} \displaystyle\int\! dE_e'\,
\frac{d\Phi_\nu}{dE_\nu}\,L^{-2}\,
\sigma(E_\nu)\,\varepsilon(E_e')\,R(E_e,E_e')\,P(L/E_\nu)\ ,
\label{eq:Etheo}
\end{equation}
%...........................................................................
where $E_e'$ and $E_e$ are the {\em true\/} and {\em measured\/}
positron energy respectively, $E_\nu$ is the neutrino energy,
$d\Phi_\nu/dE_\nu$ is the neutrino energy spectrum, $\sigma$
is the neutrino cross section, $\varepsilon$
is the detector efficiency, $R$ is the resolution function, 
$P$ is the oscillation probability, and $\cal N$ is a normalization factor.

	The  Michel  energy spectrum $d\Phi_\nu/dE_\nu$ of $\bar\nu_\mu$'s
from $\mu^+$ decay at rest is well known  (see, e.g., Fig.~6 in \cite{At96b}). 
The total flux varies  approximately with the inverse square of the distance  
$L$ (see Sec.~II~B of \cite{At96a}).

	The neutrino cross section $\sigma$ for the reaction (\ref{eq:POS}) 
is also well known \cite{Ll72}. The  neutrino energy and the (true) positron 
energy $E_e'$ are tightly related,  
$E_\nu = E_e' + 1.8 {\rm\ MeV}+\delta(\theta_e)$, where $\delta$ is a small 
(and often neglected) kinematical correction depending on the  positron 
scattering angle $\theta_e$. We have  applied a fixed correction corresponding 
to the average scattering angle $\cos\theta_e\simeq0.2$ (see Sec.~VI~C of 
\cite{At96a}).

	The efficiency  $\varepsilon(E_e')$ for detecting a positron with 
true  energy $E_e'$ can be recovered from Fig.~9 in \cite{At96a}. It varies 
from  $40\%$ to $90\%$ above the analysis threshold 
($ E_e \geq 20{\rm \ MeV}$).

	The difference between the measured and true positron energy is 
well approximated by a Gaussian energy resolution function
$R(E_e,\,E_e')$  with one-sigma width equal to 7.7\% at 52.8 MeV, and scaling 
as $1/E_e'$ for other energies (see Sec. VI~D of \cite{At96b}).

	Finally, the normalization factor $\cal N$ is fixed by imposing 
that for 100\% transmutation ($P=1$) the event excess $E_{\rm theo}$ 
(integrated over the detector length and over the energy range 
$20 {\rm\ MeV}\leq E_e \leq 60 {\rm\ MeV}$) be equal to 16670
events, as reported in Table~IV of \cite{At96a}.

	We have checked that, for $P=1$, the shape of $dE_{\rm theo}/dE_e$
derived from Eq.~(\ref{eq:Etheo})  compares very well with the corresponding
LSND simulation (Fig.~7 in \cite{At96a}). Moreover, our estimated value 
of $E_{\rm theo}$, integrated over the energy subrange 
$36 {\rm\ MeV} \leq E_e \leq 60 {\rm\ MeV}$,  coincides with the LSND 
quoted value (12500 events, see Table~III of \cite{At96a})
within 0.5\%. These two nontrivial checks add confidence
in our calculation of the event excess due to neutrino oscillations 
through Eq.~(\ref{eq:Etheo}).

	Concerning the background $B$,  the relevant distributions
$dB/dE_e$ and $dB/dL$  are not  reported  explicitly in the LSND publications 
\cite{At96b,At96a}.  We have derived them indirectly by subtraction, $B=S-E$. 
More precisely, Figs.~30(a) and 30(d) in \cite{At96a} show (as solid lines) 
the distributions $dS_{\rm theo}/dE$ and $dS_{\rm theo}/dL$  of the signal 
expected for the oscillation case 
$(\Delta m^2_\nu,\,\sin^2 2\theta)=(19 {\rm\ eV}^2,\,0.006)$.
We calculate $dE_{\rm theo}/dE$ and $dE_{\rm theo}/dL$
from Eq.~(\ref{eq:Etheo}) for the same oscillation  parameters and then obtain
the background distributions by subtraction, $dB=dS_{\rm theo}-dE_{\rm theo}$.

	Figs.~\ref{F:6} and \ref{F:7} show, respectively,  the distributions
$dB/dE_e$ and $dB/dL$ derived in this way (dashed histograms). 
The integrated background is $B=1701$ events. To guide the eye, we also 
show in  Figs.~\ref{F:6} and \ref{F:7} the 
corresponding distributions of $S_{\rm exp}$  (dots with error bars) 
and of $S_{\rm theo}$ [for  $(m^2,\,\sin^2 2\theta)=(19 {\rm\ eV}^2,\,0.006)$], 
as taken from Figs.~30(a) and 30(d) of \cite{At96a} respectively.
The integrated  signal is $S_{\rm exp}=1763$ events, corresponding to 
$S_{\rm exp}-B=62$  excess events.

	With the above ingredients we calculate, for a given oscillation 
probability function $P$, the  $E$ and $L$ distributions of the expected 
signal $S_{\rm theo}$ and the associated likelihood [Eq.~(\ref{eq:Fact})],
from which the confidence contours of Fig.~\ref{F:1} have been obtained.

	When used in combination with other neutrino oscillation
data, the LSND likelihood is transformed in a $\chi^2$ through the relation 
\cite{PD96} $\chi^2=-\ln {\cal L}/2$.

%%%%%%%%%%%%%%%%%%%%%%%%%%%%%%%%%%%%%%%%%%%%%%%%%%%%%%%%%%%%%%%%%%%%%%%%%%%%
\subsection{Variations in the analysis}

	The ``standard'' analysis of the LSND data described in the previous
section (and used throughout this work) makes use of the $E_e$ and $L$
distributions of the data (Figs.~\ref{F:6} and \ref{F:7}) through a
maximum likelihood analysis.

	In order to test the stability of the ``standard'' LSND limits shown 
in Fig.~\ref{F:1}, we have performed a few alternative  analyses of the LSND 
data that gradually deviate from the standard one. The results are shown in Fig.~\ref{F:8}.

	Fig.~\ref{F:8}(a) shows the 90 and 99 \% C.L.\ limits obtained
by dropping the likelihood factor ${\cal L}_L$ in Eq.~(\ref{eq:Fact}).
The limits are only slightly relaxed with respect to Fig.~\ref{F:1}.
In fact, the fit is basically driven by the excess of events in a few bins
of the energy distribution (Fig.~\ref{F:6}), while the path length
distribution (Fig.~\ref{F:7}) is not really discriminating within the 
statistical error bars. For the sake of simplicity, only the dominant $E_e$ 
distribution is used in all panels of Fig.~\ref{F:8}.

	Fig.~\ref{F:8}(b) shows the LSND limits from a ``true''
$\chi^2$ analysis of the $E_e$ distribution (i.e., the canonical definition 
of $\chi^2$ is used and not $-\ln {\cal L}/2$). 
Only statistical errors are included. A comparison with Fig.~\ref{F:8}(a) 
shows a slight relaxation of the C.L.\ contours for small mixing (i.e., for 
lower rates), as expected from the  Gaussian approximation of  a Poisson 
distribution which is implied  by the $\chi^2$ method.

	Fig.~\ref{F:8}(c) shows the results of the same $\chi^2$ analysis of
Fig.~\ref{F:8}(b) with the addition of a plausible 10\% uncertainty
in the background normalization \cite{At96a}. The correlation of this
error between any two bins is taken equal to 1. The LSND oscillation
limits are somewhat relaxed, and the data became even compatible with
no oscillations at about $ 99\%$ C.L.

	In Fig.~\ref{F:8}(d) the same analysis of Fig.~\ref{F:8}(c) is 
repeated by dividing the energy distribution in 5 bins instead of 20. The 
rationale for this exercise is the erratic position of bins where there is a
significant event excess (see  Fig.~\ref{F:6}), that might be a symptom
of statistical fluctuations.
Such fluctuations  should be somewhat flattened by grouping
bins. In fact, the LSND limits appear to be slightly relaxed as compared
with  Fig.~\ref{F:8}(c). A more evident effect is obtained by dividing
the $E_e$ distribution in just two bins, as shown in Fig~\ref{F:8}(e).

	Finally, Fig.~\ref{F:8}(f) shows the LSND limits obtained when the
$E_e$ distribution is fully integrated, i.e., when only the total number
of events is used. Since the total  event excess (62 events) is smaller than 
the systematic uncertainty of the  background (10\% of 1701 events),  no 
significant indication in favor of neutrino oscillations is obtained, and 
the analysis gives only {\em exclusion\/} contours. Therefore, the preference 
for nonzero values of the neutrino
oscillation parameters in Figs.\ref{F:8}(a)--(e) appears to be driven by 
the information contained in the positron energy spectrum.

	In conclusion, the exercises described in this section show that,
as far as the published information is used, the ``standard'' LSND limits of 
Fig.~\ref{F:1} appear to be dominated by the detailed energy distribution of 
the observed event excess. Different ways of treating this distribution may 
lead to significant changes in the C.L.\ contours. In particular, the 
indications in favor of neutrino oscillations are increasingly weakened by 
gradually integrating the energy spectrum information.

%%%%%%%%%%%%%%%%%%%%%%%%%%%%%%%%%%%%%%%%%%%%%%%%%%%%%%%%%%%%%%%%%%%%%%%%%%%%
\section{Reanalysis of the KARMEN data}
\label{app:KAR}
%%%%%%%%%%%%%%%%%%%%%%%%%%%%%%%%%%%%%%%%%%%%%%%%%%%%%%%%%%%%%%%%%%%%%%%%%%%%

	The Karlsruhe Rutherford Medium Energy Neutrino (KARMEN) experiment
\cite{Kl96,Kl97,Ei96} is being performed at the pulsed spallation neutron 
facility ISIS of the Rutherford Appleton Laboratory. A search is made for 
$\nu_e$ and $\bar\nu_e$ appearance from $\nu_\mu$ and $\bar\nu_\mu$
produced in $\pi^+$ and $\mu^+$ decay at rest. Therefore, the 
energy spectrum of the neutrino source,  $d\Phi_\nu/dE$, is the same as in 
the LSND experiment. The KARMEN detector,  a 56 ton liquid scintillation 
calorimeter, is located at a distance $L=17.5\pm 1.75$~m (front-end range) 
from the source. The pulsed time  structure of the beam can be exploited 
to achieve a strong suppression of the background.

	At present, the  number of  candidate events is consistent with the 
estimated background for  both oscillation channels, 
$\bar\nu_\mu\leftrightarrow\bar\nu_e$  and $\nu_\mu\leftrightarrow\nu_e$. 
These negative results can be used to constrain the neutrino mass and mixing 
parameters. Our statistical analysis of the constraints for the two
oscillation channels and their combination is detailed below.

%%%%%%%%%%%%%%%%%%%%%%%%%%%%%%%%%%%%%%%%%%%%%%%%%%%%%%%%%%%%%%%%%%%%%%%%%%%%
\subsection{Channel $\bar\nu_\mu\leftrightarrow\bar\nu_e$}

	For this channel we use three basic inputs \cite{Kl97}:
(1)  the  number of events (prompt positrons) 
observed in excess of the expected background, 
$N_{\rm exp}\pm\sigma_N=-0.4^{+6.9}_{-5.3}$;
(2)  the  number of events expected for 100\% transmutation,
$N_{\rm max}=3038$; and 
(3) the  positron energy spectrum $S(E_{e^+})$ (Fig.~6 of \cite{Ei96})
expected for 100\% transmutation. 
The above inputs refer to the default energy window
$20\leq E_{e^+}\leq 52$ MeV. The approximate energy relation 
$E_{\bar\nu_e}\simeq E_{e^+} + 1.8$ MeV is adopted.

	The expected (theoretical) number of positrons for a generic 
oscillation probability $P=P(L/E_{\bar\nu_e})$ is then given by
%...........................................................................
\begin{equation}
N_{\rm theo} = N_{\rm max}\frac
{\displaystyle \int_{L-D/2}^{L+D/2}dL\,L^{-2}\int dE_{e^+}\,S(E_{e^+})\,
P(L/E_{\bar\nu_e})}
{\displaystyle \int_{L-D/2}^{L+D/2}dL\,L^{-2}\int dE_{e^+}\,S(E_{e^+})}\ ,
\end{equation}
%...........................................................................
where $D$ is  the detector length (3.5 m).

	The probability $\epsilon$ that a difference as large as $\Delta=
(N_{\rm theo}-N_{\rm exp})/\sigma_N$ be a statistical fluctuation is obtained 
by integrating a one-sided Gaussian ($N_{\rm theo}$ being nonnegative),
%...........................................................................
\begin{equation}
1-\epsilon = \sqrt{\frac{2}{\pi}} \int^\Delta_0 \! dx \, \exp(-x^2/2)\ .
\end{equation}
%...........................................................................

	As a check, we draw  in Fig.~\ref{F:9}  the
exclusion curve (dotted line)  corresponding to $1-\epsilon=90\%$ in the usual
two-flavor mass-mixing plane.  This curve matches well the 90\% C.L.
contour reported by the KARMEN collaboration in \cite{Kl97}
for the same oscillation channel.

	When a combination with other oscillation data is performed 
in a global $\chi^2$ analysis,  we use a ``fake'' $\chi^2$ statistic 
determined by probability inversion $[\chi^2 = \chi^2 (\epsilon)]$,
yielding for any assigned C.L.\ the same exclusion curves.

%%%%%%%%%%%%%%%%%%%%%%%%%%%%%%%%%%%%%%%%%%%%%%%%%%%%%%%%%%%%%%%%%%%%%%%%%%%%
\subsection{Channel $\nu_\mu\leftrightarrow\nu_e$}

	In this channel the energy spectrum is monochromatic 
($E_{\nu_\mu}=29.8$ MeV).  We use three basic inputs \cite{Kl97}: 
(1)  the number of candidate events, $N=1$; 
(2) the estimated background $N_B$ and its 
1-sigma uncertainty, $N_B\pm \sigma_B=2\pm 0.3$; and 
(3) the number of events expected for 100\% transmutation, $N_{\rm max}=154$.

	For a generic oscillation probability, 
the expected number of events is  then calculated as 
%...........................................................................
\begin{equation}
N_{\rm theo} = N_{\rm max}\frac 
{\displaystyle \int_{L-D/2}^{L+D/2}dL\,L^{-2}
P(L/E_{\nu_\mu})}
{\displaystyle \int_{L-D/2}^{L+D/2}dL\,L^{-2}}\ .
\end{equation}
%...........................................................................

	The confidence level $1-\epsilon$  associated to $N_{\rm theo}$ can 
be calculated by using the statistics appropriate to a Poisson process with 
background \cite{PD96},  with an allowance for
statistical fluctuations of the background itself \cite{He83}:
%...........................................................................
\begin{equation}
1-\epsilon = 1-\frac
{\displaystyle\int^\infty_0 dn_B\,f(n_B)\,e^{-(n_B+N_{\rm theo})}
\sum_{n=0}^N (n_B+N_{\rm theo})^n/n!}
{\displaystyle\int^\infty_0 dn_B\,f(n_B)\,e^{-n_B}
\sum_{n=0}^N n_B^n/n!}\ ,
\end{equation}
%...........................................................................
where
%...........................................................................
\begin{equation}
f(n_B)=\frac{1}{\sqrt{2\pi}\sigma_B}
\exp{\left[-\frac{1}{2}\left(\frac{n_B-N_B}{\sigma_B}\right)^2\right]}\ .
\end{equation}
%...........................................................................

	The curve corresponding to $1-\epsilon=90\%$ in the usual
mass-mixing plane is shown in  Fig.~\ref{F:9} as a thin, solid line. It 
reproduces in detail the 90\% C.L. contour reported by the KARMEN 
collaboration in \cite{Kl97}.

	As for the antineutrino channel,  we use a ``fake'' $\chi^2$ 
statistic determined by probability inversion $[\chi^2 = \chi^2 (\epsilon)]$ 
when a combination with other data is performed. In particular,
we combine the two  KARMEN oscillation channels,
$\nu_\mu\leftrightarrow\nu_e$ and  $\bar\nu_\mu\leftrightarrow\bar\nu_e$, by 
summing the  corresponding $\chi^2$'s. The curve corresponding to a variation
of 4.61 in the total $\chi^2$ (90\% C.L.\ for two degrees of
freedom) is shown in Fig.~\ref{F:9} as a thick, solid curve
(the same curve  reported in the top left panel of
Fig.~\ref{F:2}). It can be seen from a comparison of the three
curves in Fig.~\ref{F:9} that the channel  
$\bar\nu_\mu\leftrightarrow\bar\nu_e$  dominates the KARMEN exclusion region.

%%%%%%%%%%%%%%%%%%%%%%%%%%%%%%%%%%%%%%%%%%%%%%%%%%%%%%%%%%%%%%%%%%%%%%%%%%%%

%%%%%%%%%%%%%%%%%%%%%%%%%%%%%%%%%%%%%%%%%%%%%%%%%%%%%%%%%%%%%%%%%%%%%%%%%%%%
%			F I G U R E S
%%%%%%%%%%%%%%%%%%%%%%%%%%%%%%%%%%%%%%%%%%%%%%%%%%%%%%%%%%%%%%%%%%%%%%%%%%%%
\begin{figure}
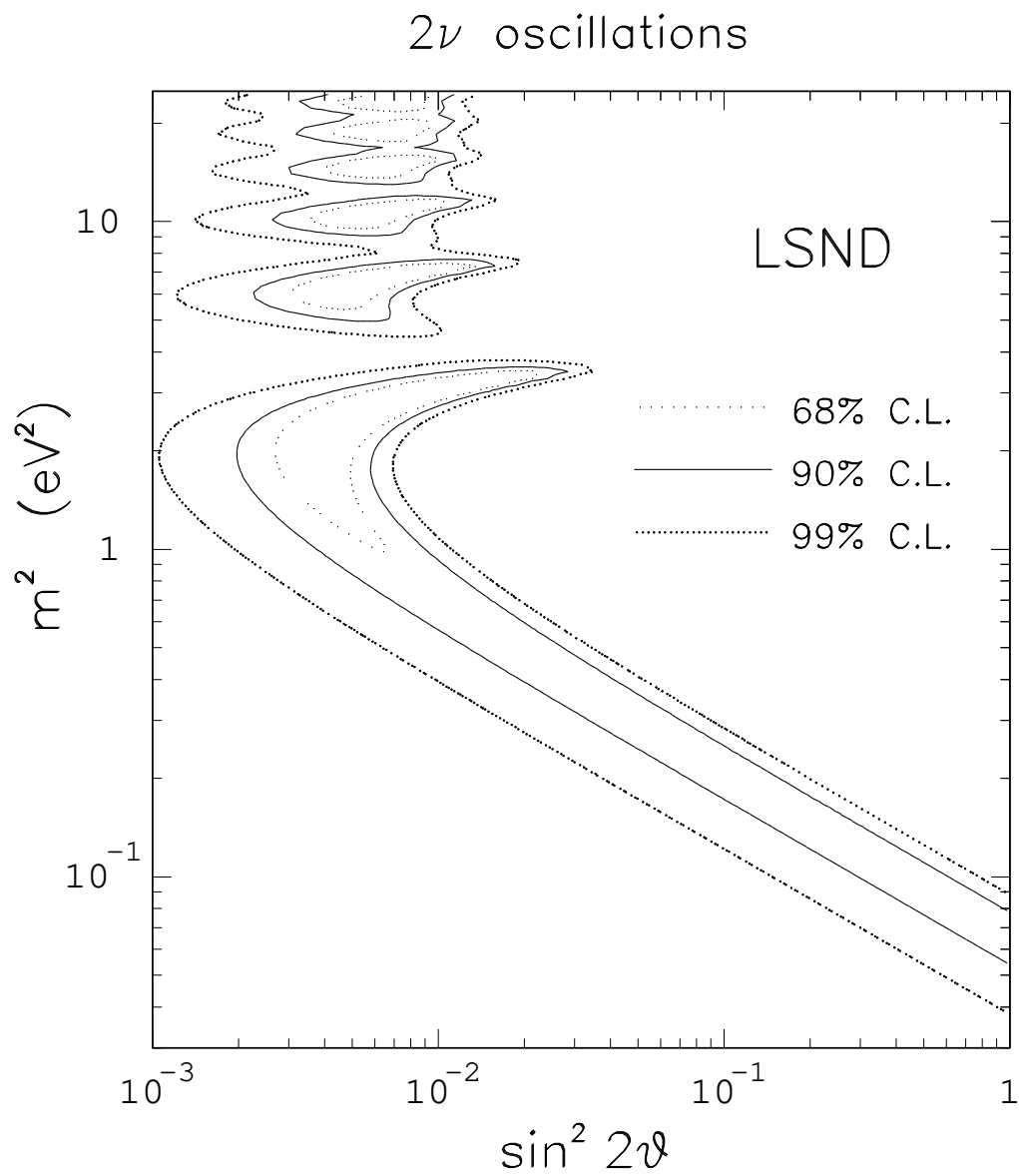

\caption{	Region of the $2\nu$ oscillation parameters preferred
		by the LSND data (our reanalysis)
		at 68, 90, and 99 \% C.L.\ ($N_{\rm DF}=2$).}
\label{F:1}
\end{figure}
\vskip-1mm
%...........................................................................
\begin{figure}
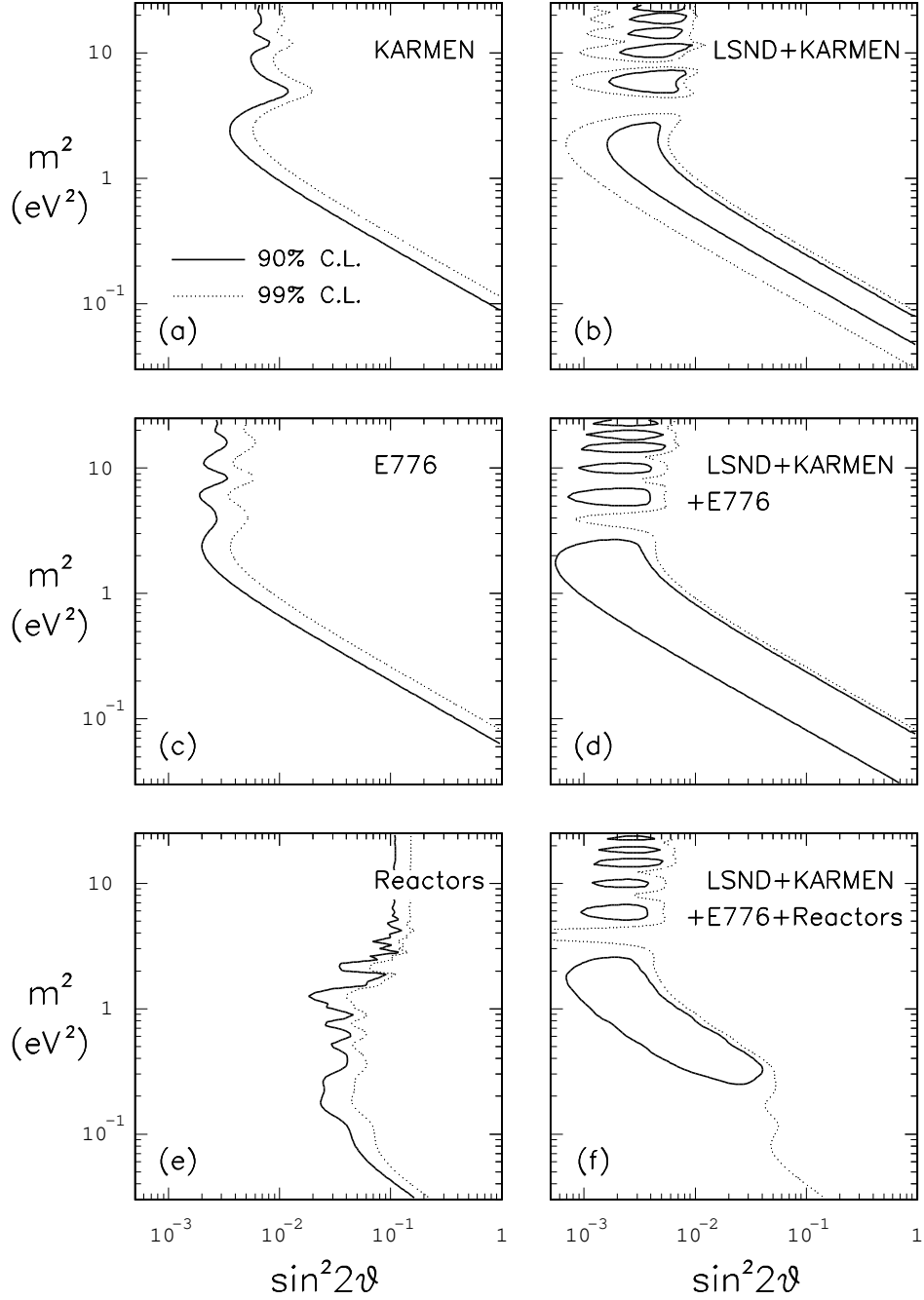

\caption{	Laboratory oscillation data in two flavors (our reanalysis).
		Left panels: contours of regions excluded at 90 (solid) 
		and 99 \% (dashed) C.L.\ by the accelerator experiments KARMEN
		and E776, and by reactor experiments (G{\"o}sgen, Bugey,
		and Krasnoyarsk combined). Right panels: variations in
		the region preferred by the LSND data 
		(see Fig.~\protect\ref{F:1}) with the progressive
		addition of the KARMEN, E776, and reactor data.}
\label{F:2}
\end{figure}\vskip-1mm
%...........................................................................
\begin{figure}
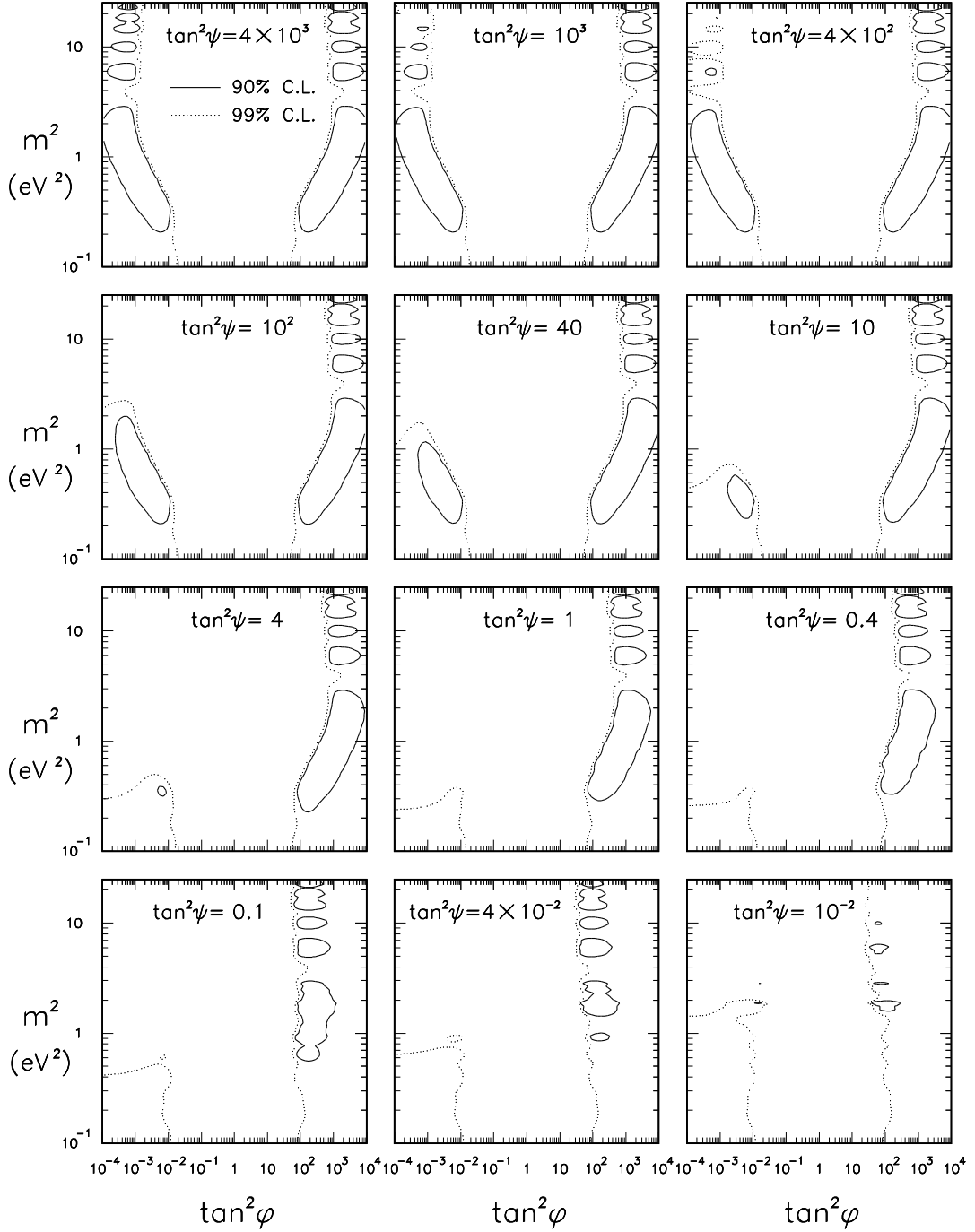

\caption{	Three-flavor analysis of the most constraining laboratory 
		oscillation experiments (LSND, KARMEN, E776, E531, CDHSW, 
		G{\"o}sgen, Bugey, and Krasnoyarsk combined). The
		preferred region in the $3\nu$ parameter space 
		$(m^2,\,\tan^2\psi,\,\tan^2\phi)$ is shown through
		twelve $(m^2,\,\tan^2\phi)$ sections at fixed,
		representative values of $\tan^2\psi$. Solid lines:
		90\% C.L.\ contours ($\Delta\chi^2=6.25$ for 
		$N_{\rm DF}=3$). Dotted lines: 99\% C.L.\ contours 
		($\Delta\chi^2=11.36$).}
\label{F:3}
\end{figure}\vskip-1mm
%...........................................................................
\begin{figure}
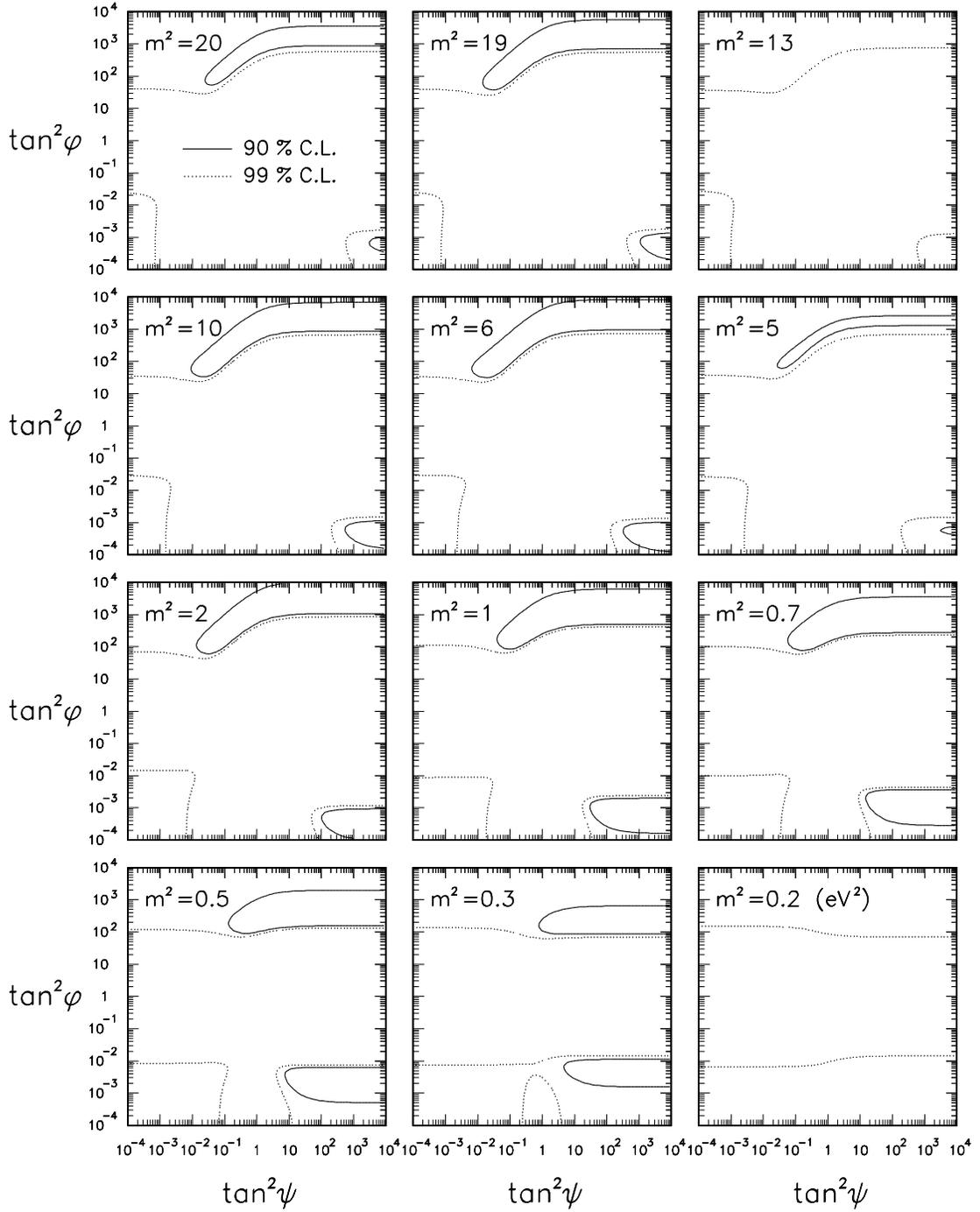

\caption{	As in Fig.~\protect\ref{F:3}, but in 
		$(\tan^2\phi,\,\tan^2\psi)$ sections at twelve 
		representative values of $m^2$.}
\label{F:4}
\end{figure}\vskip-1mm
%...........................................................................
\begin{figure}
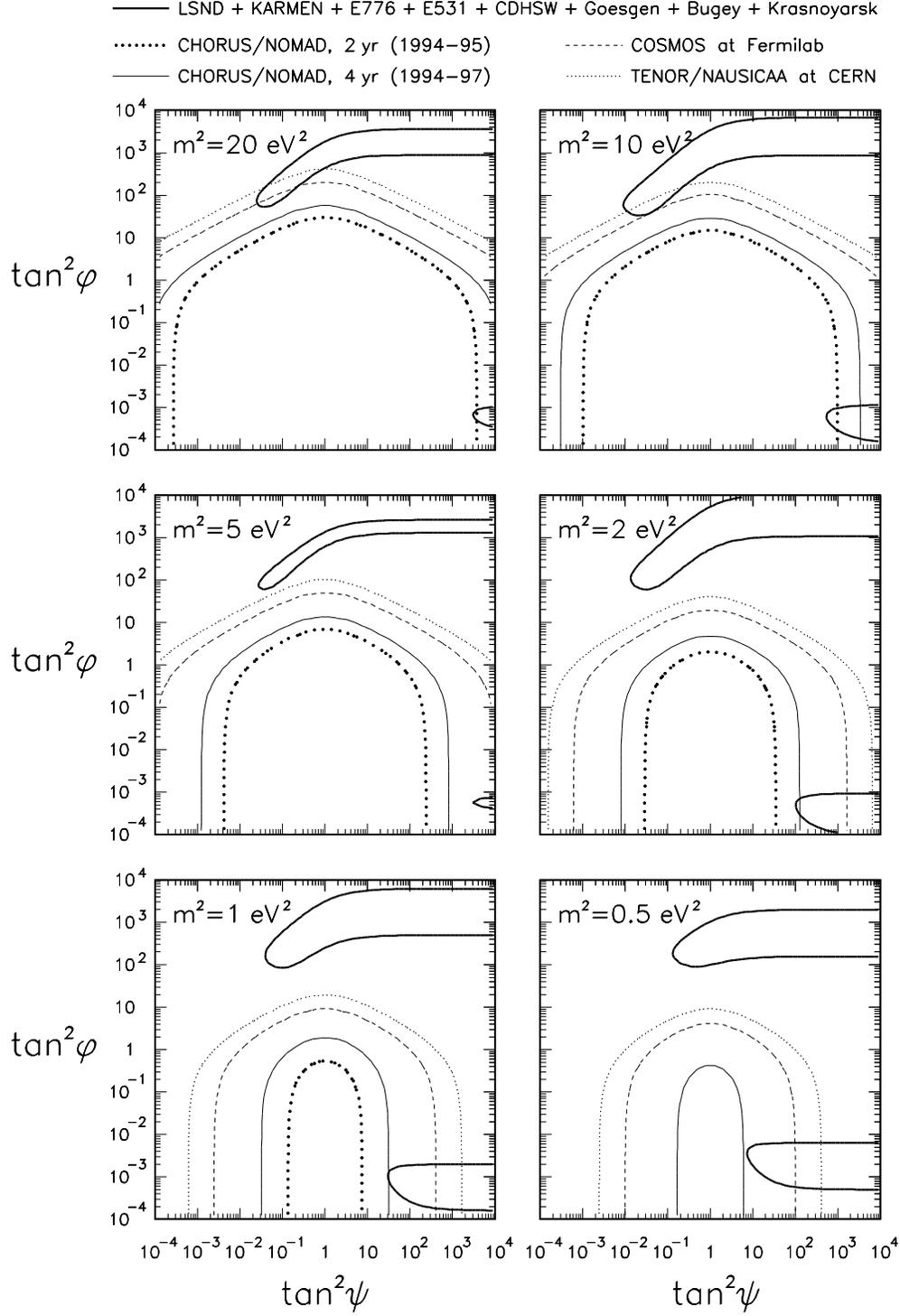

\caption{	Regions of the parameter space explorable at 90\% C.L.\
		by the following $\nu_\mu\to\nu_\tau$ experiments 
		(in order of increasing sensitivity): CHORUS or
		NOMAD in two years (thick, dotted line) and four years
		(thin, solid line), COSMOS (dashed line), and
		TENOR or NAUSICAA (thin, dotted line). These experiments can
		probe a fraction of the zone preferred 
		at 90\% C.L.\ by the combination
		of all the available data, including LSND (thick, solid line).}
\label{F:5}
\end{figure}\vskip-1mm
%...........................................................................
\begin{figure}
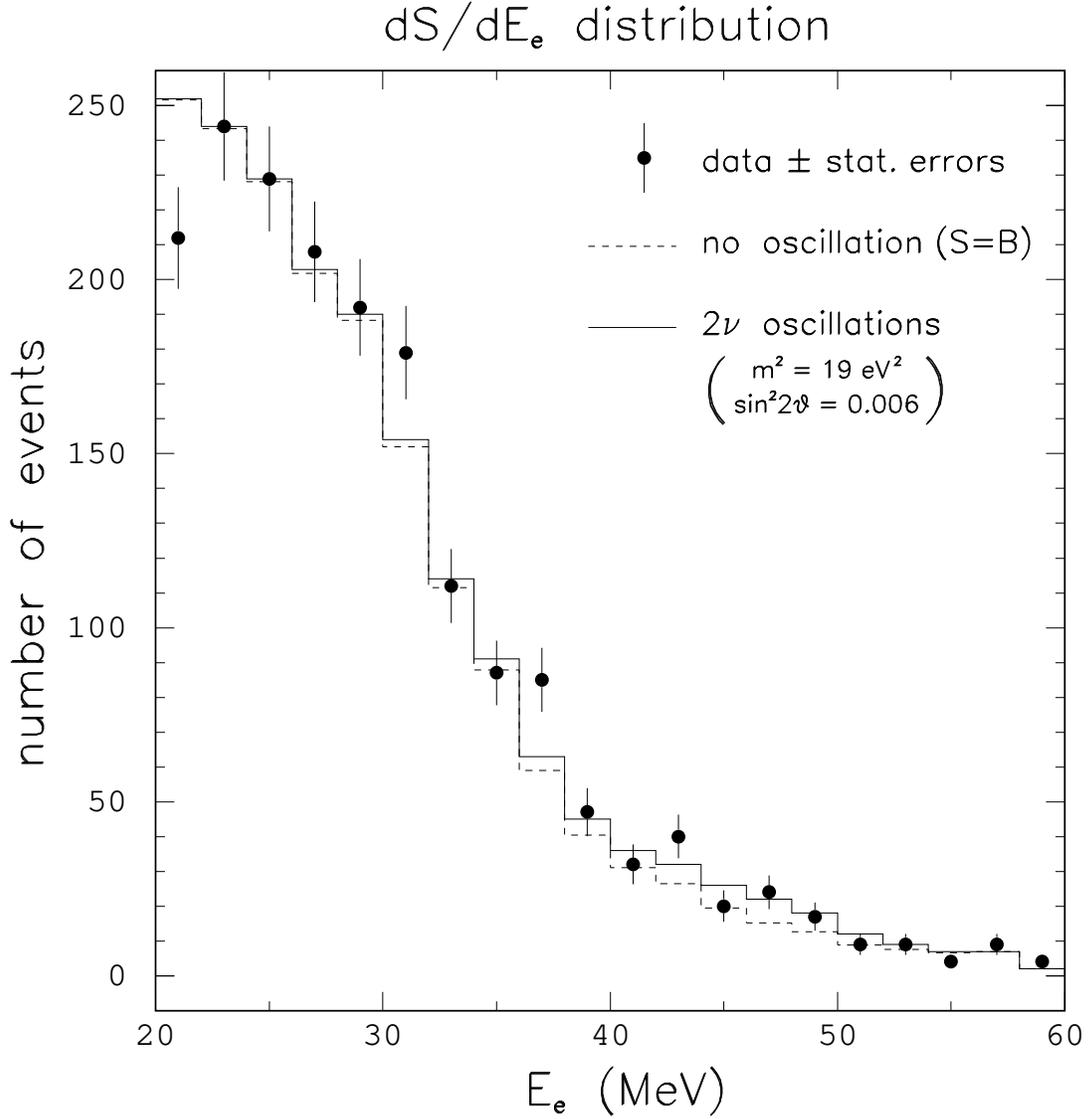

\caption{	Energy distribution $dS/dE_e$ of the LSND signal (20 bins).
		Dashed line: background component $dB/dE_e$ (our reanalysis).
		Solid line: signal $dS_{\rm theo}/dE_e$ expected for
		$(m^2,\,\sin^2 2\theta)=(19 {\rm\ eV}^2,\,0.006)$, as taken
		from Fig.~30(a) of \protect\cite{At96a}. Dots with
		statistical error bars: observed signal 
		$dS_{\rm exp}/dE_e$, from Fig.~30(a) of 
		\protect\cite{At96a}.}
\label{F:6}
\end{figure}\vskip-1mm
%...........................................................................
\begin{figure}
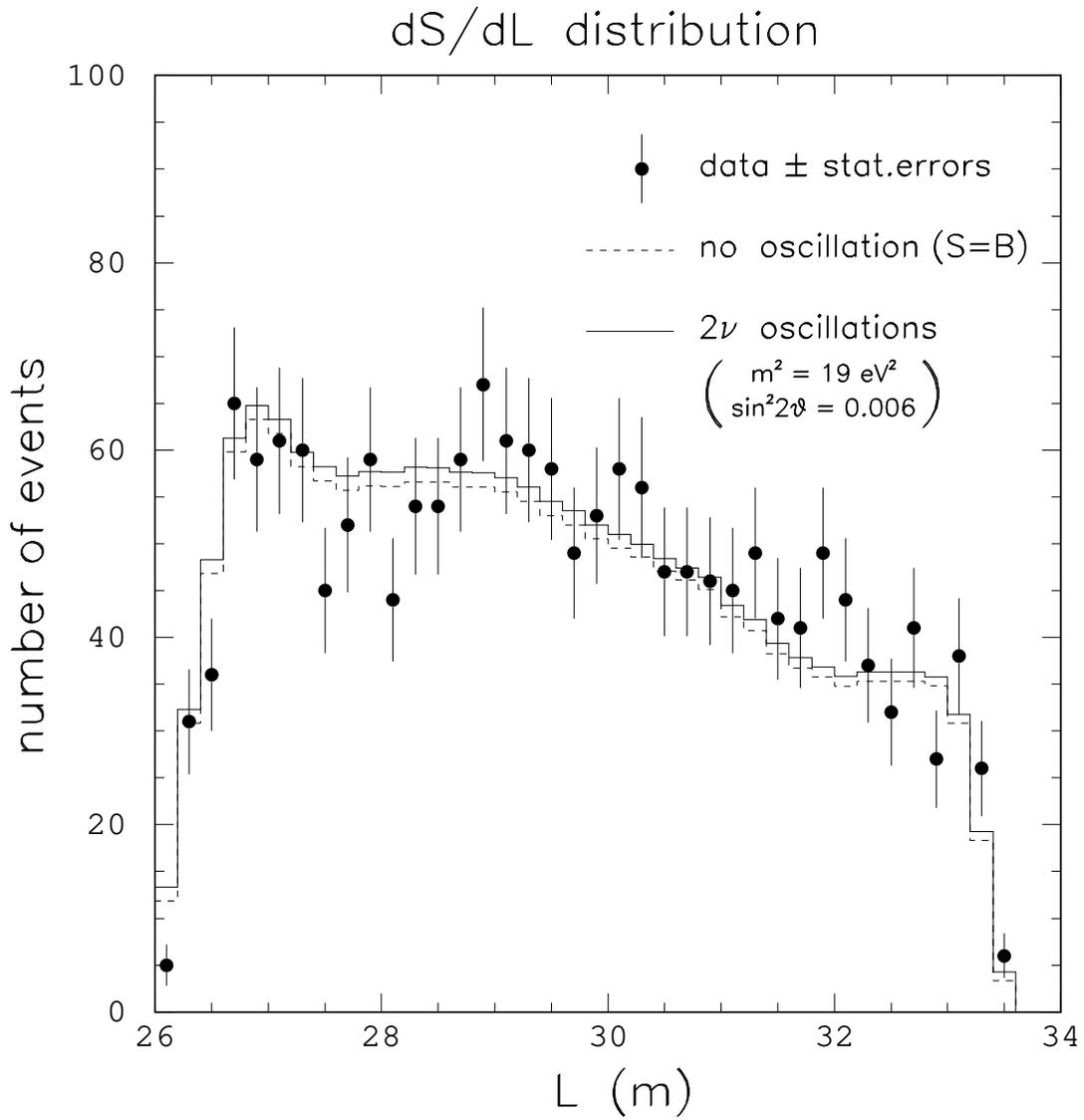

\caption{	As in Fig.~\protect\ref{F:6}, but for the path length
		distribution $dS/dL$ of the LSND signal (38 bins).}
\label{F:7}
\end{figure}\vskip-1mm
%...........................................................................
\begin{figure}
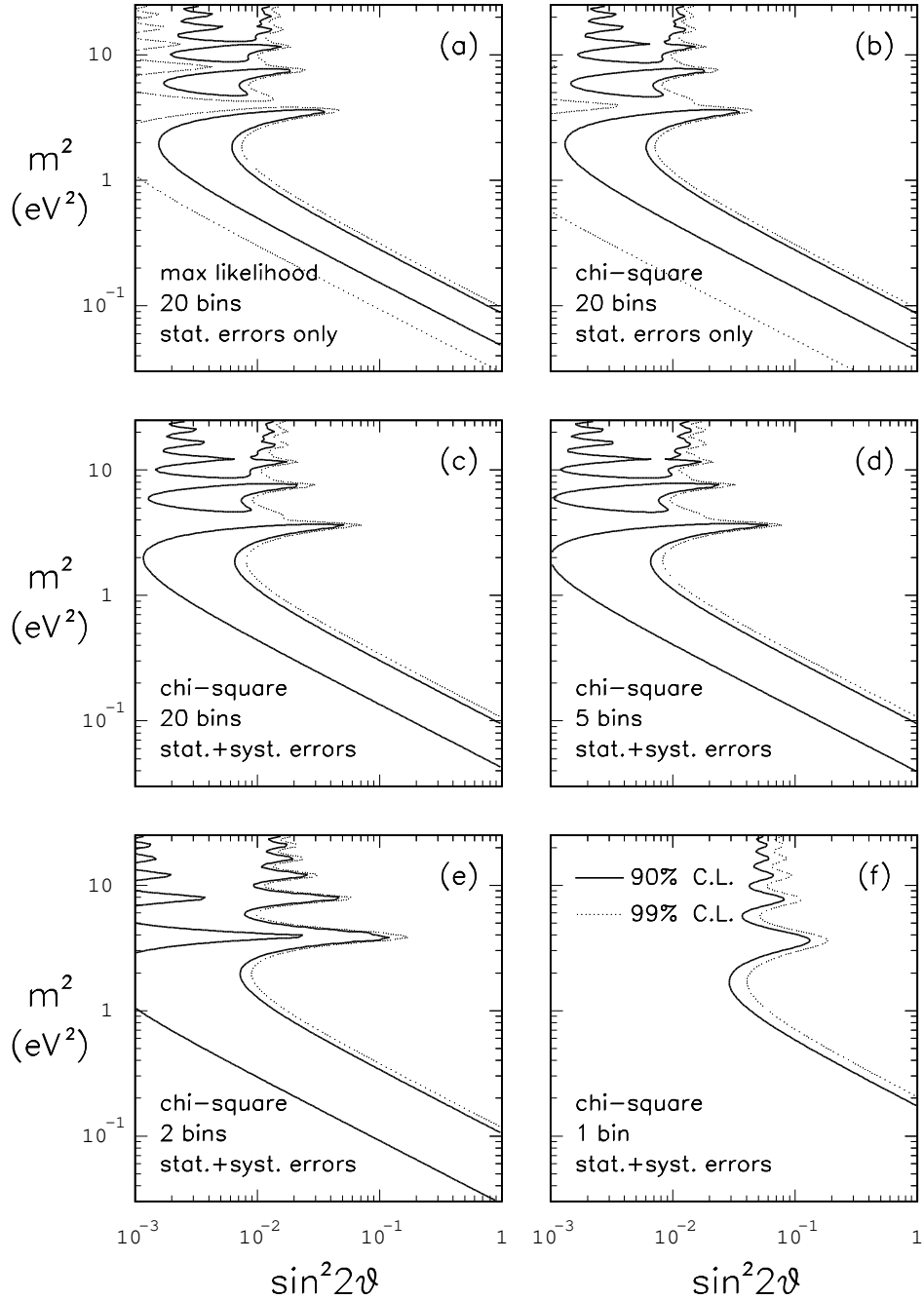

\caption{	Variations in the  LSND bounds with respect
		to the ``standard'' bounds of
		Fig.~\protect\ref{F:1}, as a result of
		of data analyses different from  
		${\cal L}_{EL}$ maximization.
		(a) Maximization of ${\cal L}_E$ only.
		(b) $\chi^2$ analysis of the energy distribution
		of the signal with statistical
		errors only. (c) $\chi^2$ analysis of the energy 
		distribution, assuming a systematic 10\% uncertainty
		in the overall background normalization. (d) As in (c),
		but dividing the energy distribution in 5 bins.
		(e) As in (c), but dividing the energy distribution
		in 2 bins. (f) As in (c), but integrating the total
		signal ($=1$ bin). See the text for details.}
\label{F:8}
\end{figure}\vskip-1mm
%...........................................................................
\begin{figure}
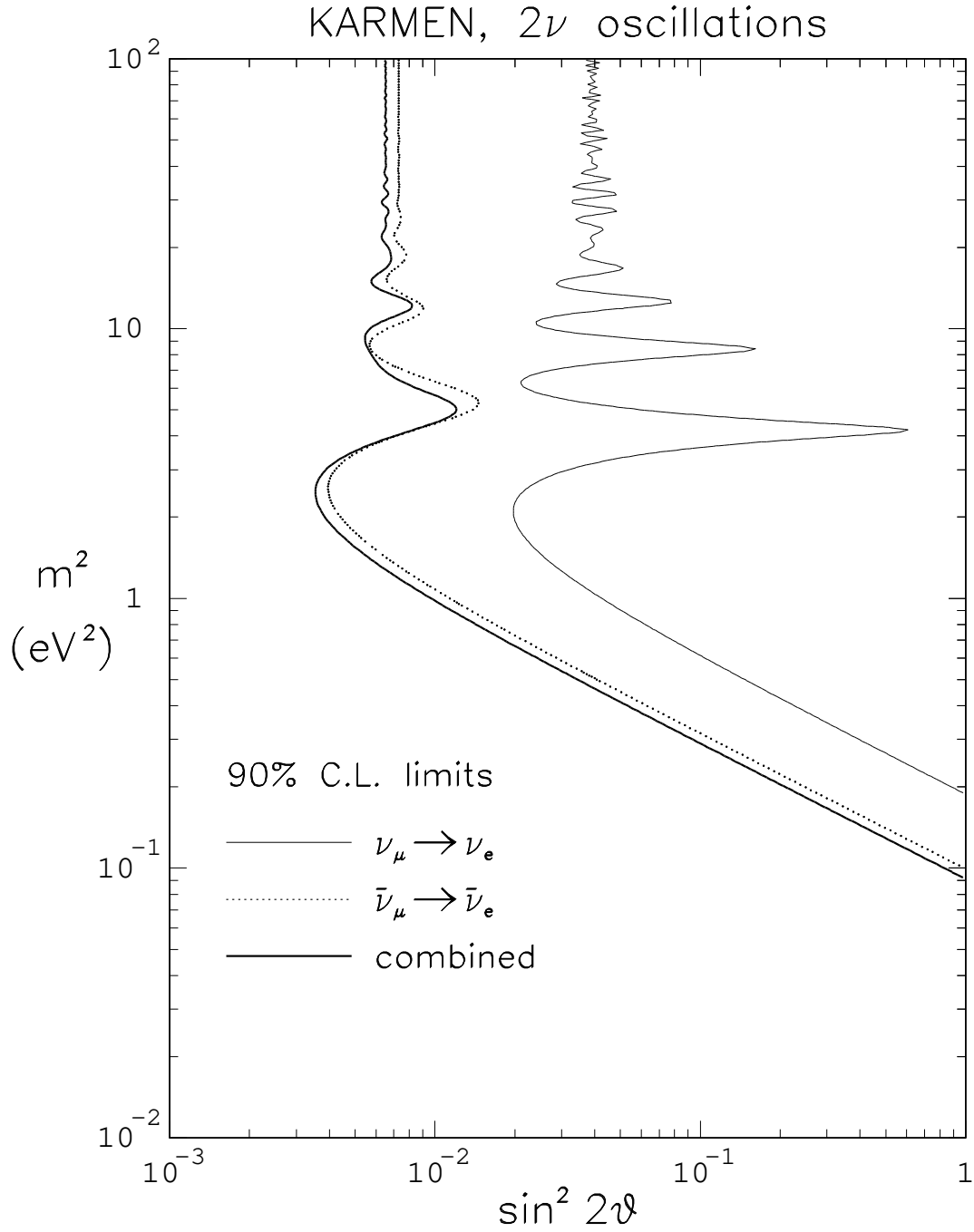

\caption{	Results of our reanalysis of the KARMEN data for the neutrino
		and antineutrino channels and their combination.
		Contours are drawn at 90\% C.L.\ ($\Delta\chi^2=4.61$
		for $N_{\rm DF}=2$).}
\label{F:9}
\end{figure}

%%%%%%%%%%%%%%%%%%%%%%%%%%%%%%%%%%%%%%%%%%%%%%%%%%%%%%%%%%%%%%%%%%%%%%%%%%%%
%\end{document}
%%%%%%%%%%%%%%%%%%%%%%%%%%%%%%%%%%%%%%%%%%%%%%%%%%%%%%%%%%%%%%%%%%%%%%%%%%%%

%%%%%%%%%%%%%%%%%%%%%%%%%%%%%%%%%%%%%%%%%%%%%%%%%%%%%%%%%%%%%%%%%%%%%%%%%%%%
%%%%%%%%%%%%%%%%%%%%%%%%%%%%%%%%%%%%%%%%%%%%%%%%%%%%%%%%%%%%%%%%%%%%%%%%%%%%
%%%%%%% 
%%%%%%%            THE FOLLOWING FOR AUTHOR USE ONLY.
%%%%%%%
%%%%%%%            INCLUSION OF FIGURES WITH EPSFIG.STY.
%%%%%%%
%%%%%%%%%%%%%%%%%%%%%%%%%%%%%%%%%%%%%%%%%%%%%%%%%%%%%%%%%%%%%%%%%%%%%%%%%%%%
%%%%%%%          P O S T S C R I P T       F I G U R E S 
%%%%%%%
%%%%%%%   memo:  1) add epsfig in the \documentstyle
%%%%%%%          2) and move this part befor \end{document} 
%%%%%%%		 3) remove previous figure captions
%%%%%%%          4) include the following \newcommand:
%%----------------------------------------------------------------------------
\newcommand{\InsertFigure}[2]{\newpage\begin{center}\mbox{%
\epsfig{bbllx=1.4truecm,bblly=1.3truecm,bburx=19.5truecm,bbury=26.5truecm,%
height=21.truecm,figure=#1}}\end{center}\vspace*{-1.85truecm}%
\parbox[t]{\hsize}{\small\baselineskip=0.5truecm\hskip0.5truecm #2}}
%----------------------------------------------------------------------------
%%%%%%%%%%%%%%%%%%%%%%%%%%%%%%%%%%%%%%%%%%%%%%%%%%%%%%%%%%%%%%%%%%%%%%%%%%%%%%%

%..............................................................................
\InsertFigure{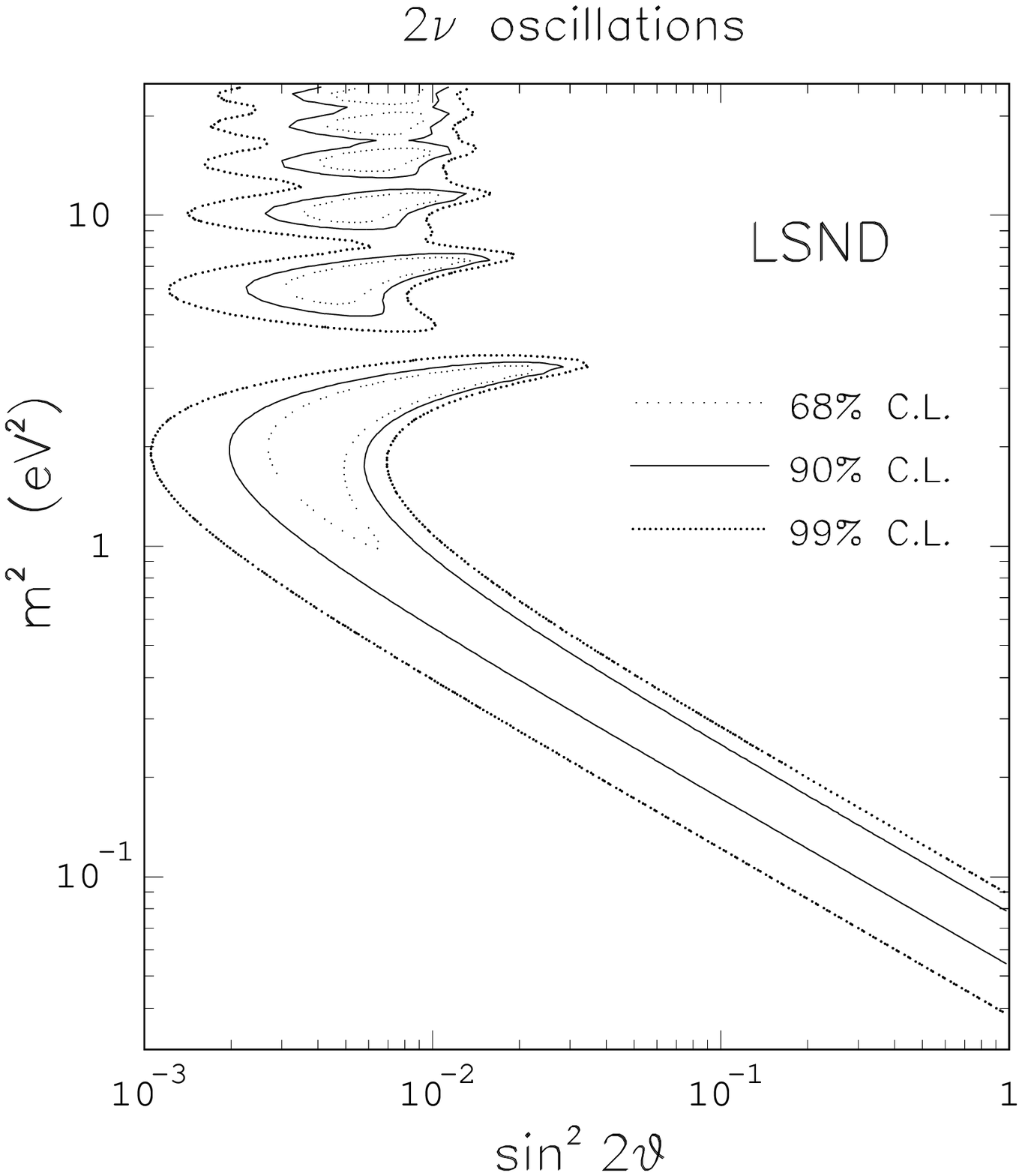}%
{FIG.~1. 	Region of the $2\nu$ oscillation parameters preferred
		by the LSND data (our reanalysis)
		at 68, 90, and 99 \% C.L.\ ($N_{\rm DF}=2$).}
%..............................................................................
\InsertFigure{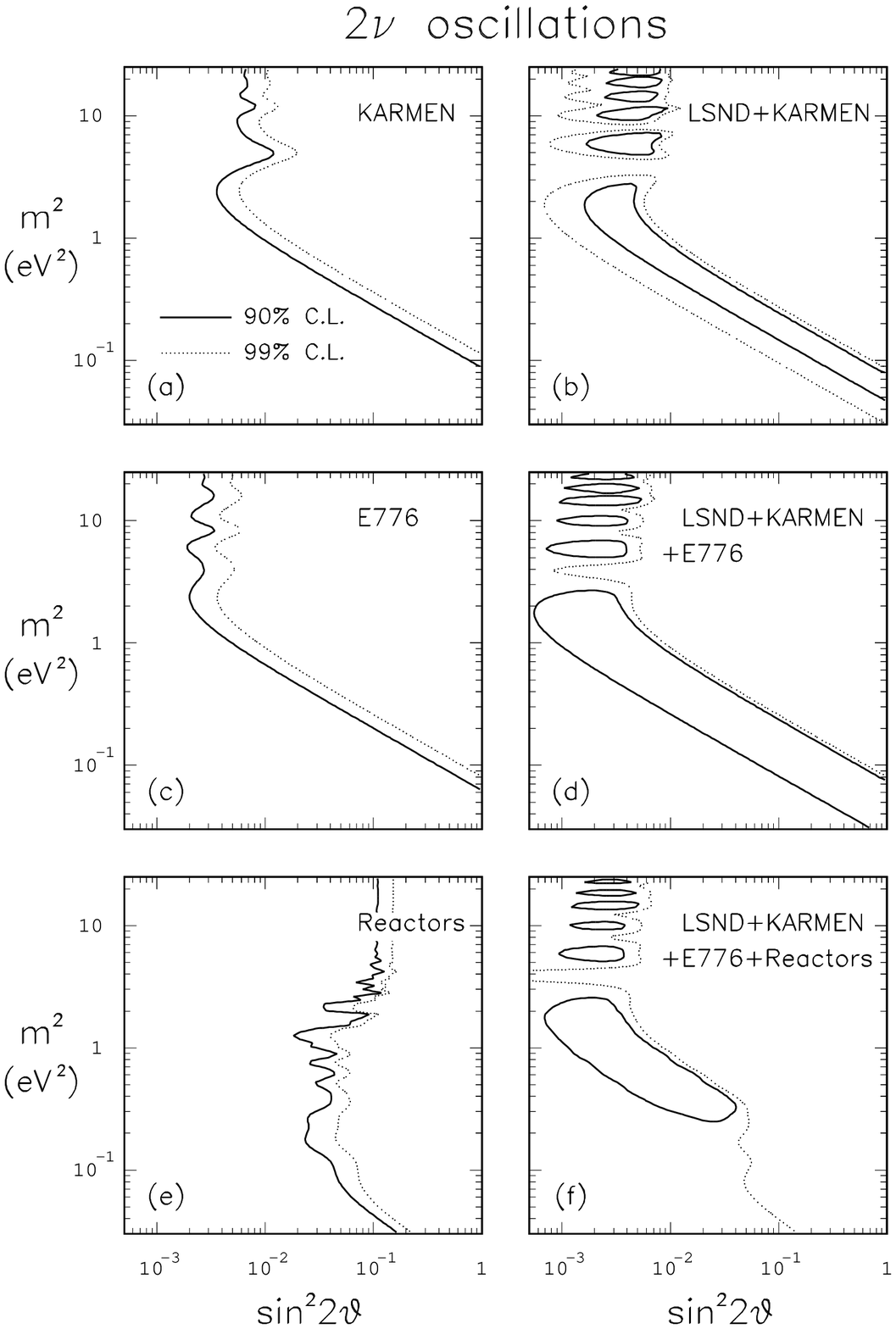}%
{FIG.~2. 	Laboratory oscillation data in two flavors (our reanalysis).
		Left panels: contours of regions excluded at 90 (solid) 
		and 99 \% (dashed) C.L.\ by the accelerator experiments KARMEN
		and E776, and by reactor experiments (G{\"o}sgen, Bugey,
		and Krasnoyarsk combined). Right panels: variations in
		the region preferred by the LSND data 
		(see Fig.~\protect\ref{F:1}) with the progressive
		addition of the KARMEN, E776, and reactor data.}
%..............................................................................
\InsertFigure{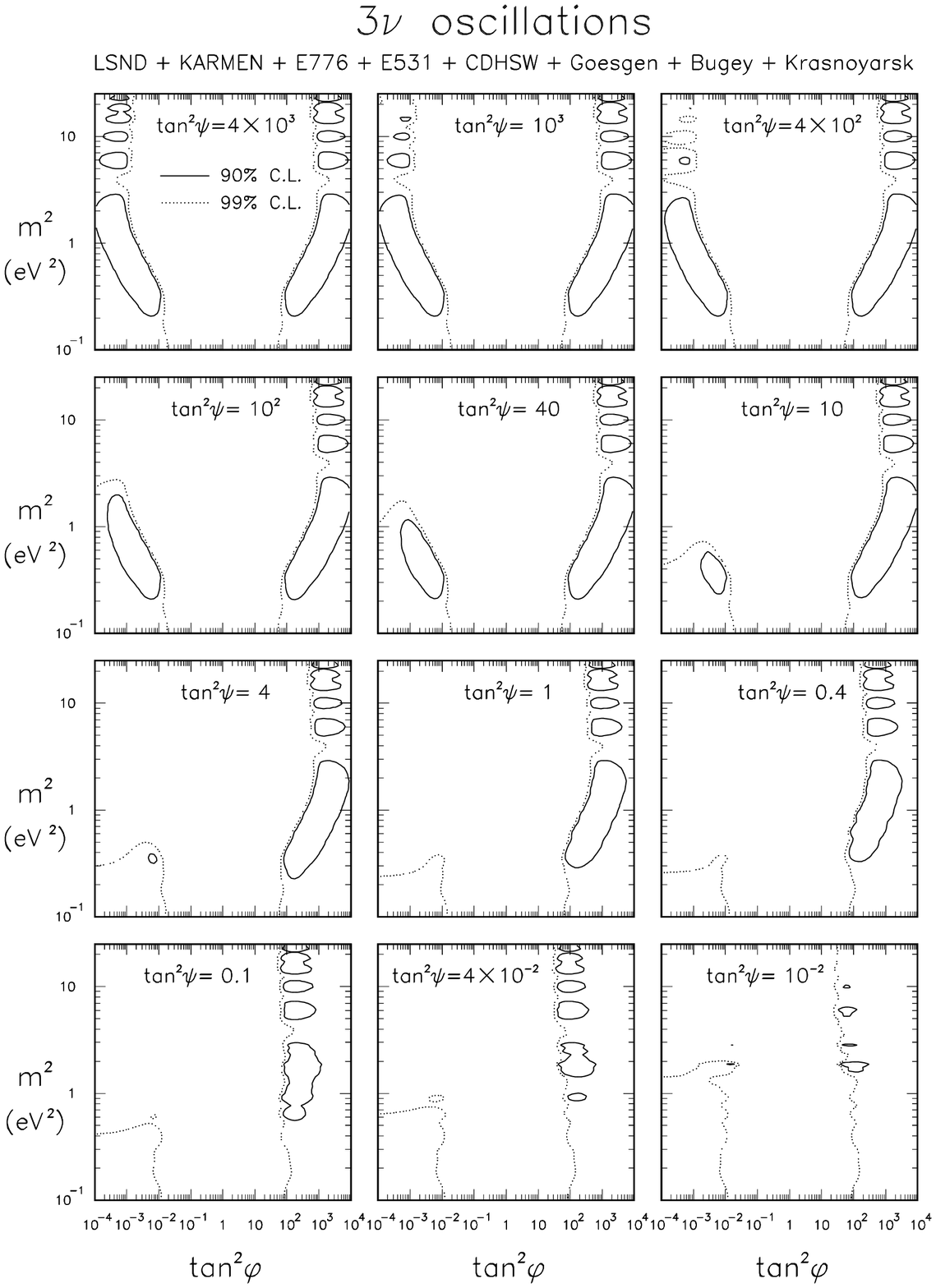}%
{FIG.~3. 	Three-flavor analysis of the most constraining laboratory 
		oscillation experiments (LSND, KARMEN, E776, E531, CDHSW, 
		G{\"o}sgen, Bugey, and Krasnoyarsk combined). The
		preferred region in the $3\nu$ parameter space 
		$(m^2,\,\tan^2\psi,\,\tan^2\phi)$ is shown through
		twelve $(m^2,\,\tan^2\phi)$ sections at fixed,
		representative values of $\tan^2\psi$. Solid lines:
		90\% C.L.\ contours ($\Delta\chi^2=6.25$ for 
		$N_{\rm DF}=3$). Dotted lines: 99\% C.L.\ contours 
		($\Delta\chi^2=11.36$).}
%..............................................................................
\InsertFigure{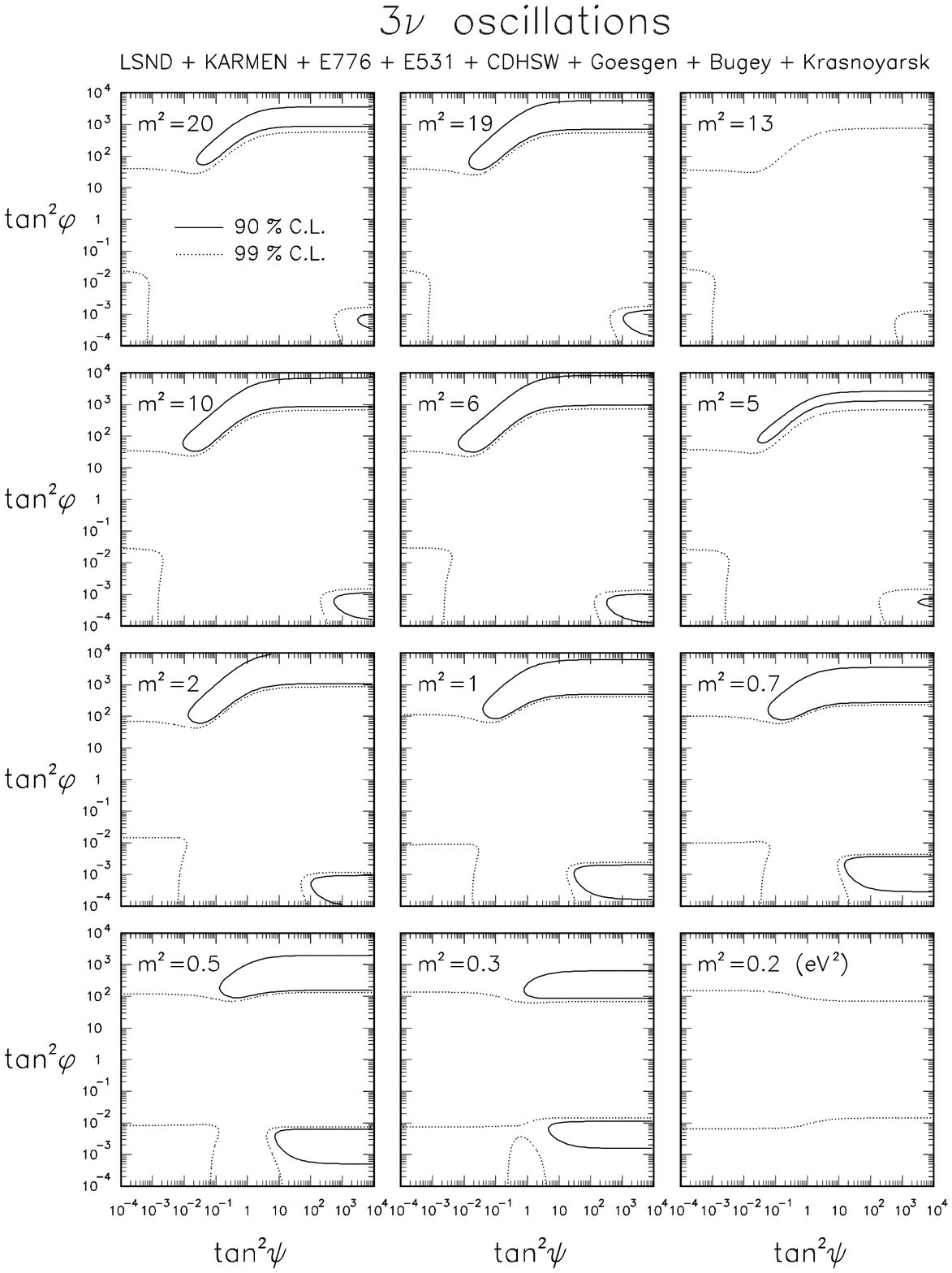}%
{FIG.~4. 	As in Fig.~\protect\ref{F:3}, but in 
		$(\tan^2\phi,\,\tan^2\psi)$ sections at twelve 
		representative values of $m^2$.}
%..............................................................................
\InsertFigure{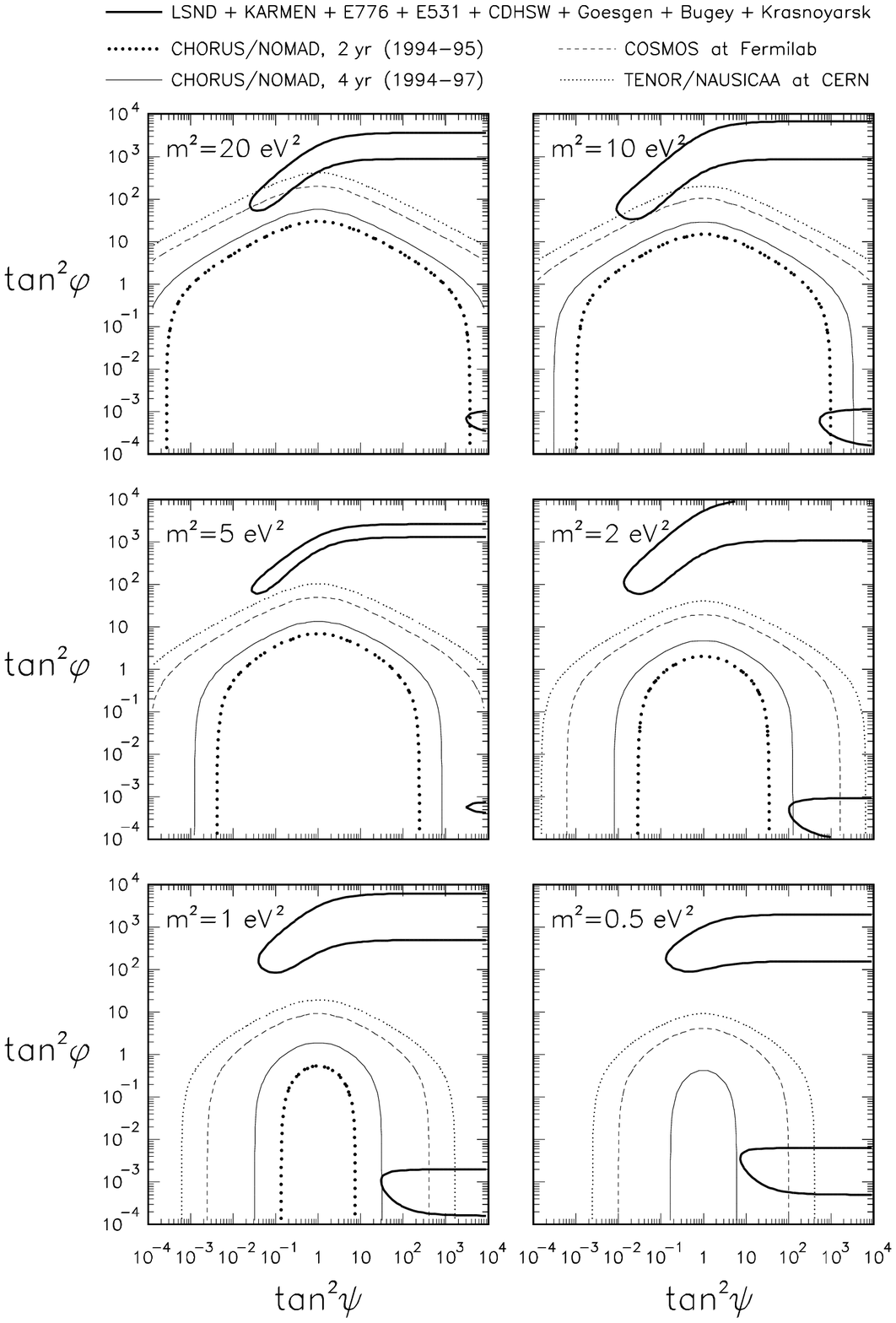}%
{FIG.~5. 	Regions of the parameter space explorable at 90\% C.L.\
		by the following $\nu_\mu\to\nu_\tau$ experiments 
		(in order of increasing sensitivity): CHORUS or
		NOMAD in two years (thick, dotted line) and four years
		(thin, solid line), COSMOS (dashed line), and
		TENOR or NAUSICAA (thin, dotted line). These experiments can
		probe a fraction of the zone preferred 
		at 90\% C.L.\ by the combination
		of all the available data, including LSND (thick, solid line).}
%..............................................................................
\InsertFigure{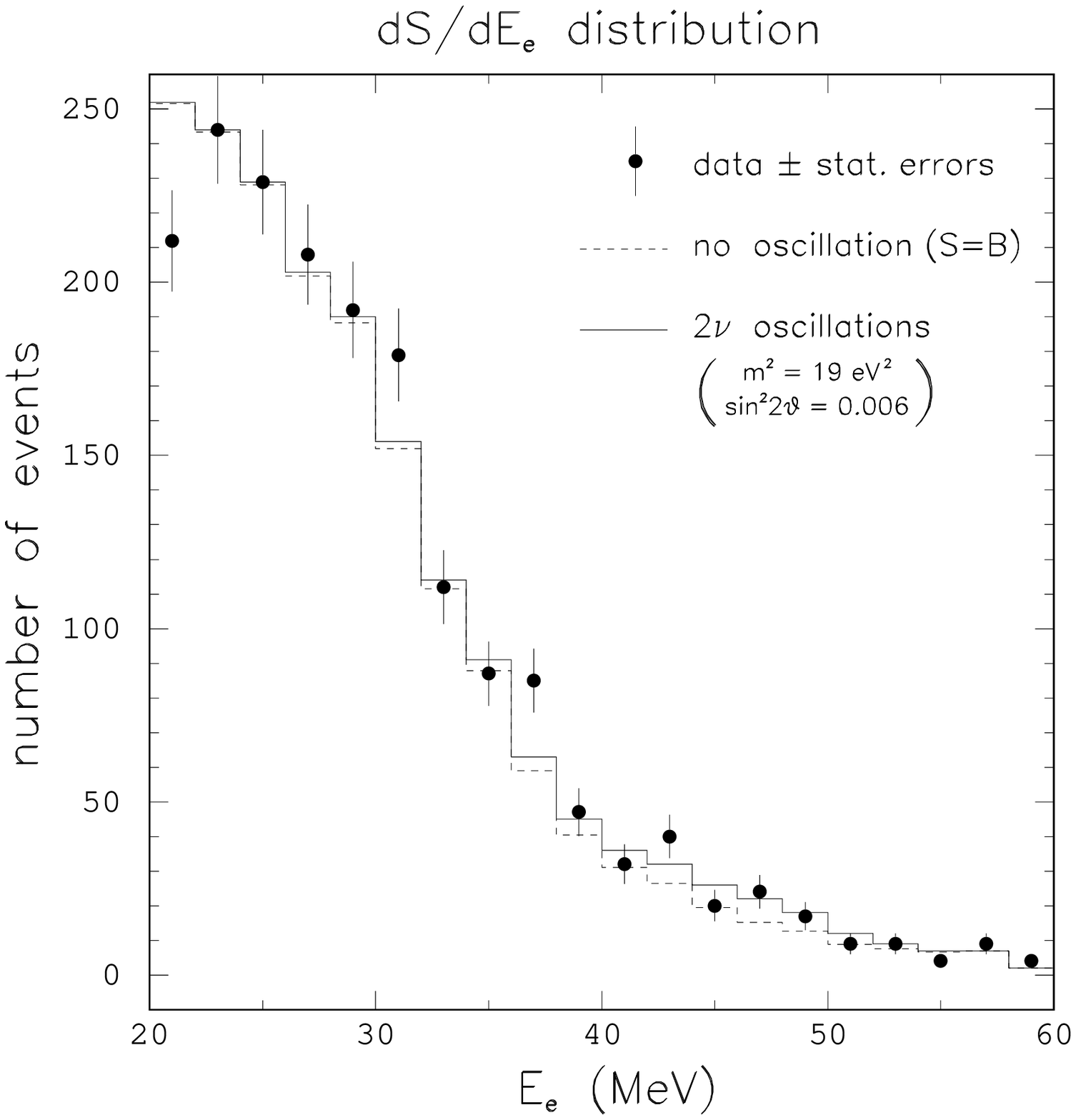}%
{FIG.~6. 	Energy distribution $dS/dE_e$ of the LSND signal (20 bins).
		Dashed line: background component $dB/dE_e$ (our reanalysis).
		Solid line: signal $dS_{\rm theo}/dE_e$ expected for
		$(m^2,\,\sin^2 2\theta)=(19 {\rm\ eV}^2,\,0.006)$, as taken
		from Fig.~30(a) of \protect\cite{At96a}. Dots with
		statistical error bars: observed signal 
		$dS_{\rm exp}/dE_e$, from Fig.~30(a) of 
		\protect\cite{At96a}.}
%..............................................................................
\InsertFigure{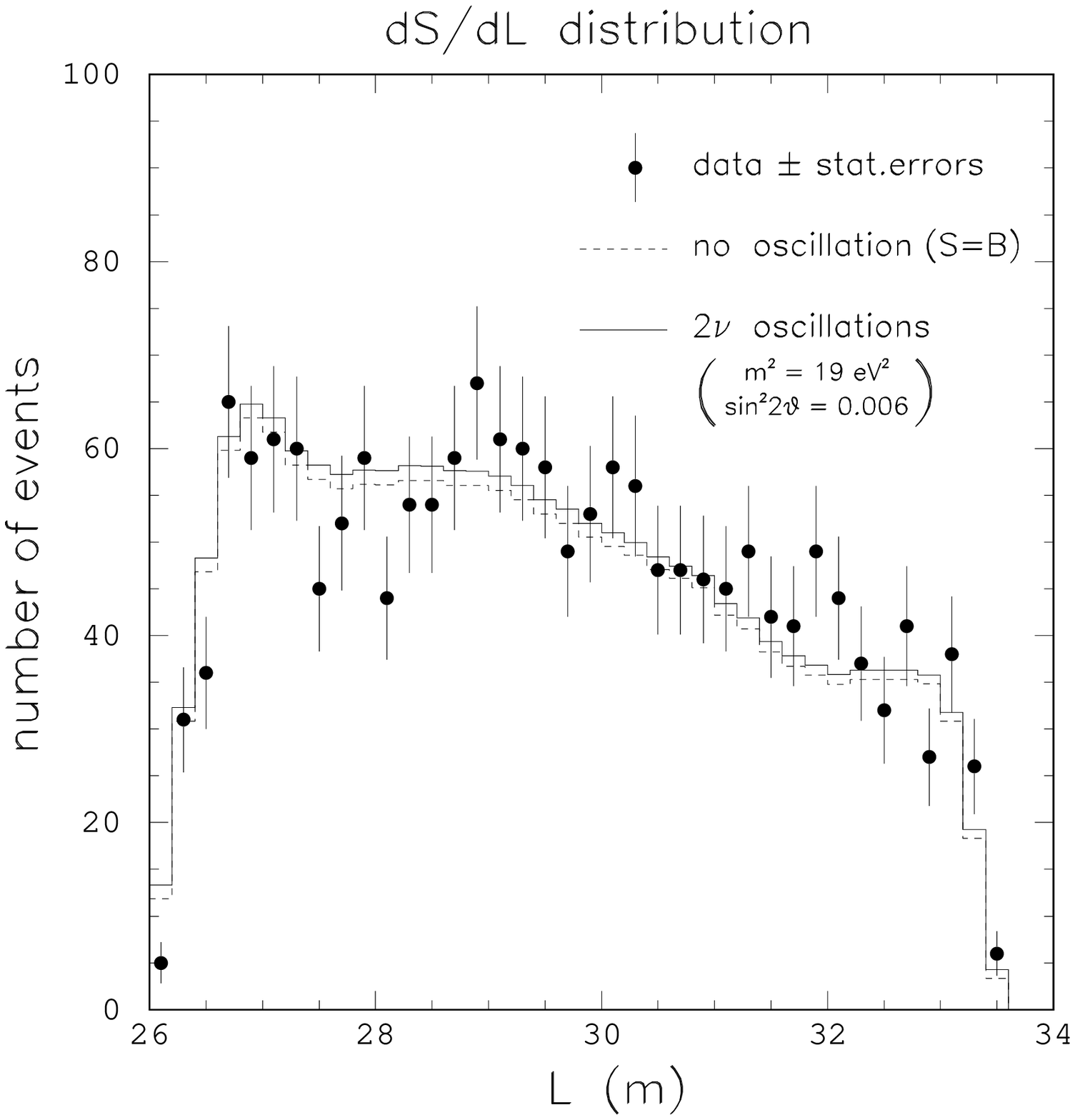}%
{FIG.~7. 	As in Fig.~\protect\ref{F:6}, but for the path length
		distribution $dS/dL$ of the LSND signal (38 bins).}
%..............................................................................
\InsertFigure{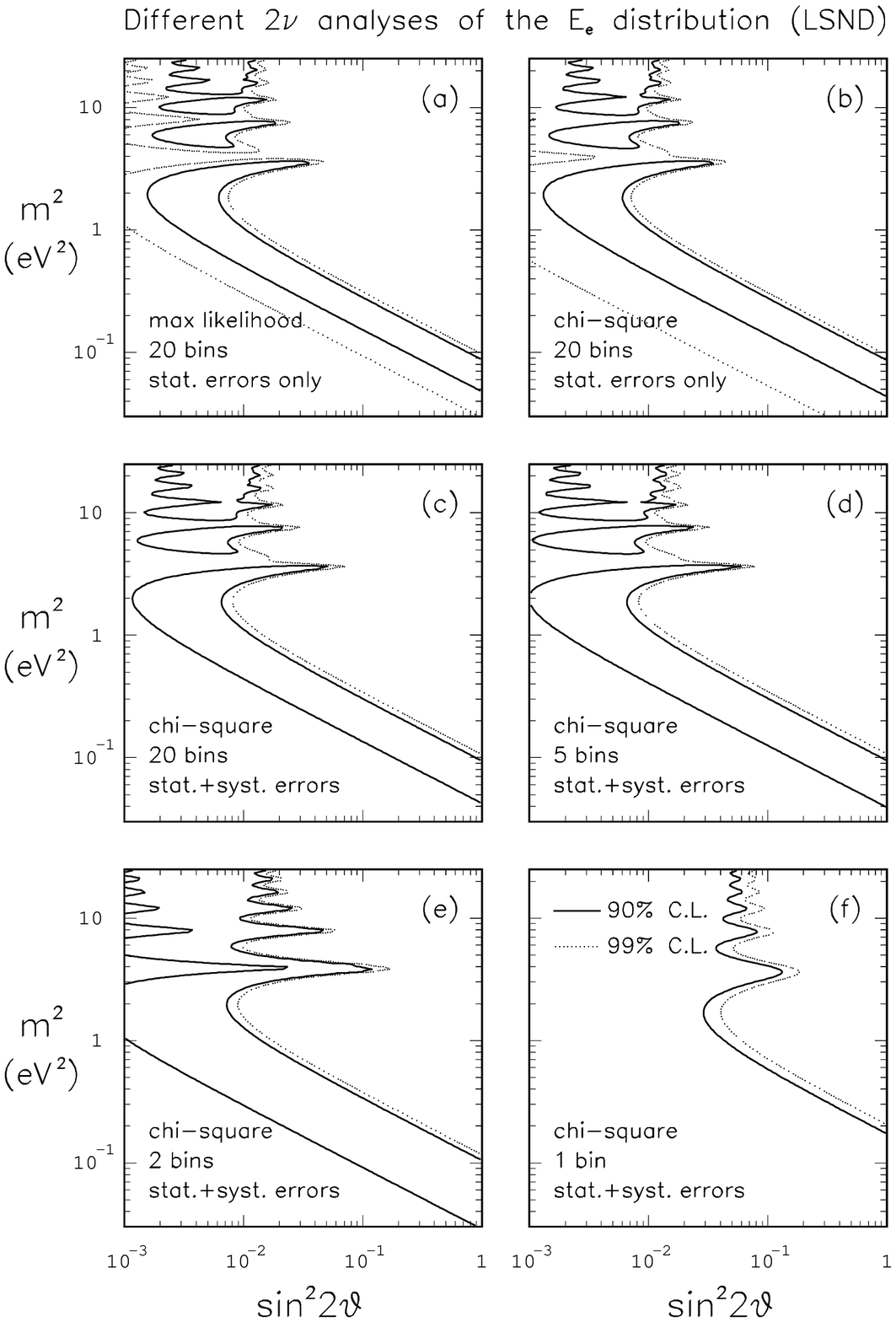}%
{FIG.~8. 	Variations in the  LSND bounds with respect
		to the ``standard'' bounds of
		Fig.~\protect\ref{F:1}, as a result of
		of data analyses different from  
		${\cal L}_{EL}$ maximization.
		(a) Maximization of ${\cal L}_E$ only.
		(b) $\chi^2$ analysis of the energy distribution
		of the signal with statistical
		errors only. (c) $\chi^2$ analysis of the energy 
		distribution, assuming a systematic 10\% uncertainty
		in the overall background normalization. (d) As in (c),
		but dividing the energy distribution in 5 bins.
		(e) As in (c), but dividing the energy distribution
		in 2 bins. (f) As in (c), but integrating the total
		signal ($=1$ bin). See the text for details.}
%..............................................................................
\InsertFigure{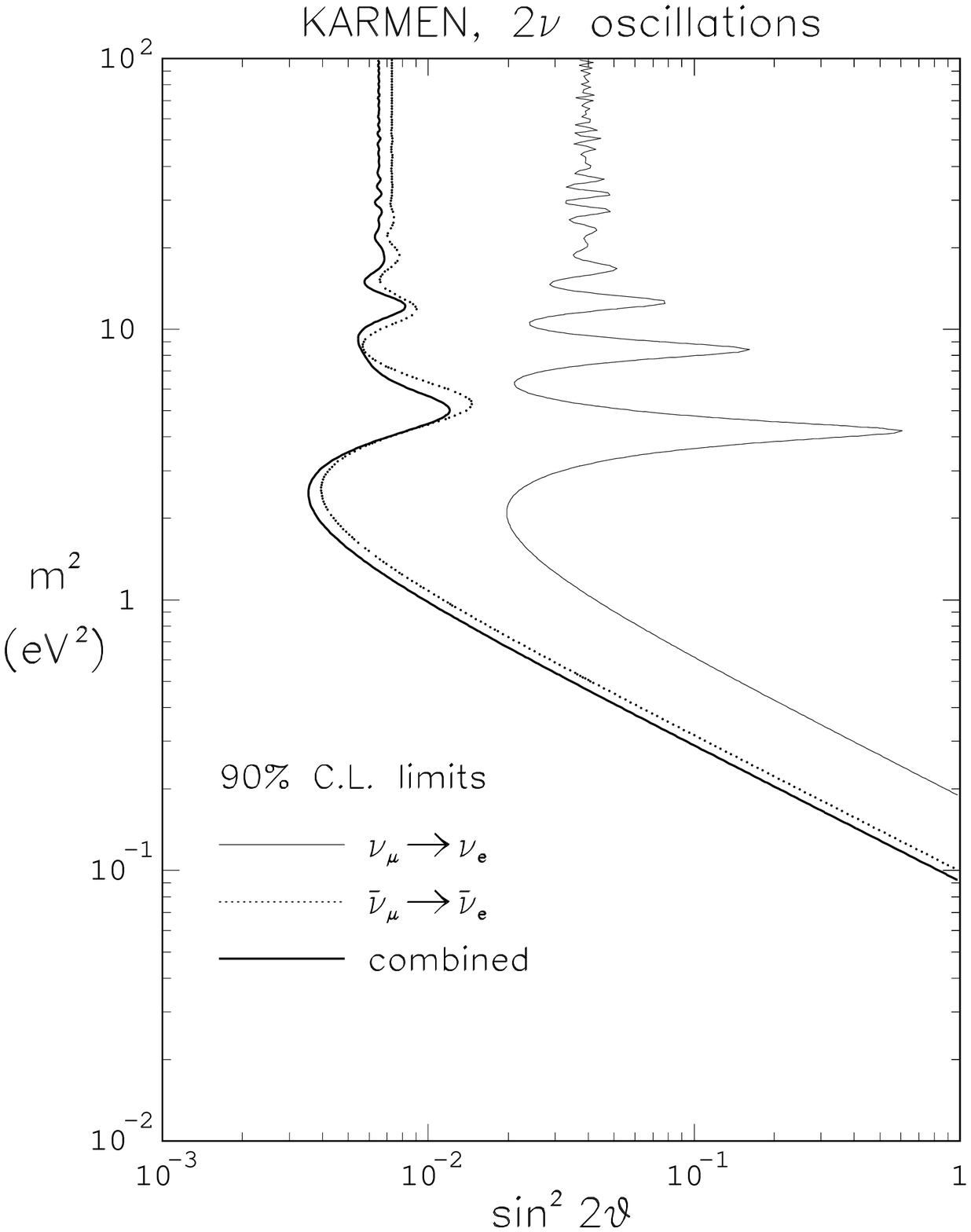}%
{FIG.~9. 	Results of our reanalysis of the KARMEN data for the neutrino
		and antineutrino channels and their combination.
		Contours are drawn at 90\% C.L.\ ($\Delta\chi^2=4.61$
		for $N_{\rm DF}=2$).}


\begin{thebibliography}{99}
\bibitem{E531}	It has been observed recently (Ref.~\protect\cite{PD96},
		p.~298) that the published number 
		of expected $\nu_e$ charged current events
		in the E531 experiment \cite{Us86} must be
		corrected by a factor of $\sim 0.48$ in the analysis of
		the $\nu_e\leftrightarrow\nu_\tau$ channel. As a consequence,
		the corresponding bounds on the mixing angle 
		become weaker (by a factor of about two). We have applied
		this correction factor in the present 
		analysis. The correction
		was not known at the time of our previous work
		\cite{Fo95}. However, the constraint
		from the E531 experiment in the $\nu_e\leftrightarrow\nu_\tau$
		are never dominant in the analysis, 
		and we have checked that the results
		of our previous global fit shown in Fig.~10 of \cite{Fo95}
		are negligibly altered by such correction.
\end{thebibliography}
\end{document}